\documentclass[prb,aps,amsmath,amssymb,amsfonts,floatfix,showpacs,twocolumn,superscriptaddress,footinbib]{revtex4}

\usepackage[colorlinks=true,citecolor=blue,urlcolor=blue,linkcolor=blue,hyperfigures=true]{hyperref}
\usepackage{graphicx}
\usepackage{amsmath,amsfonts}
\usepackage{xcolor}
\usepackage{color}

\newcommand{\up}{\uparrow}
\newcommand{\dn}{\downarrow}
\newcommand{\kv}{\ensuremath{\mathbf{k}}}

\newcommand{\av}[1]{\ensuremath{\left\langle #1 \right\rangle}}
\newcommand{\abs}[1]{\ensuremath{\left| #1 \right|}}

\renewcommand{\Re}{\operatorname{Re}}
\renewcommand{\Im}{\operatorname{Im}}

\definecolor{fmRed}{HTML}{CC3311}
\definecolor{fmBlue}{HTML}{0077BB}

\usepackage{marvosym}

\usepackage{tikz}
\usetikzlibrary{calc}
\usetikzlibrary{arrows}
\usetikzlibrary{patterns}
\usetikzlibrary{decorations.pathmorphing}
\usetikzlibrary{decorations.pathreplacing}
\usetikzlibrary{decorations.markings}
\tikzset{>=stealth}

\begin{document}

\author{Erik G. C. P. van Loon}
\affiliation{Radboud University, Institute for Molecules and Materials, NL-6525 AJ Nijmegen, The Netherlands}

\author{Friedrich Krien}
\affiliation{International School for Advanced Studies, SISSA, Trieste, Italy}
\affiliation{Institute of Theoretical Physics, University of Hamburg, 20355 Hamburg, Germany}

\author{Hartmut Hafermann}
\affiliation{Mathematical and Algorithmic Sciences Lab, Paris Research Center, Huawei Technologies France SASU, 92100 Boulogne Billancourt, France}

\author{Alexander I. Lichtenstein}
\affiliation{Institute of Theoretical Physics, University of Hamburg, 20355 Hamburg, Germany}

\author{Mikhail I. Katsnelson}
\affiliation{Radboud University, Institute for Molecules and Materials, NL-6525 AJ Nijmegen, The Netherlands}

\pacs{
}

\title{Fermion-boson vertex within Dynamical Mean-Field Theory}

\begin{abstract}
In the study of strongly interacting systems, correlations on the two-particle level are receiving more and more attention. In this work, we study a particular two-particle correlation function: the fermion-boson vertex. It describes the response of the Green's function when an external field is applied and is an important ingredient of diagrammatic extensions of Dynamical Mean-Field Theory. We provide several perspectives on this object, using Ward identities, sum rules, perturbative analysis and asymptotic relations. We then use these tools to study the vertex across the doping-driven metal-insulator transition and find a divergent imaginary part.
\end{abstract}

\maketitle

\section{Introduction}

Strongly correlated fermion systems are an area of wide physical interest that started in the 1930s and 1940s~\cite{Schubin34,Wigner34,deBoer37,Wigner38,Mott49}. The 1960s saw the introduction of the Hubbard model~\cite{Hubbard63,Gutzwiller63,Kanamori63,Hubbard64}, which has become the model for studying correlation effects such as the interaction-driven metal-insulator transition~\cite{Imada1998,Khomskii14}. Correlations are also believed to play a crucial role in high temperature superconductors~\cite{Dagotto94,Anderson97,Scalapino12}.
The advent of two-dimensional materials~\cite{Novoselov04,Novoselov05b,Novoselov05,Zhang05,Katsnelson12,Avouris17} further enhanced interest in this topic, since correlations are particularly strong in low-dimensional systems.
Recently, it has become possible to use optical lattices to create highly tunable systems of correlated fermionic atoms and molecules~\cite{Lewenstein12,Bloch12}, opening up an entirely new vista on correlation effects.  

From the theory point of view, the development of Dynamical Mean-Field Theory (DMFT)~\cite{Metzner89,Georges96} formed a major breakthrough. This theory captures all local correlation effects by means of an effective single-site model (the impurity model) subject to a self-consistently determined, dynamical field. The theory describes the transition from a metallic to an insulating state when the strength of the electron-electron interaction is increased. Today, the method forms the basis for realistic electronic structure calculations in correlated materials~\cite{Anisimov97,Lichtenstein98,Kotliar06}.

The single-particle Green's function $G$ is the central object in DMFT. The density of states can be determined directly from $G$, the insulating state is seen as an opening of the gap in the density of states at the Fermi level. The single-particle Green's function also contains information on quantities like the quasiparticle lifetime and the photoemission spectrum. However, there's more to strongly correlated systems than just $G$. After one-particle quantities, two-particle quantities are the obvious next step. These describe aspects like the spin-spin and charge-charge correlation in the material and the response to external fields.  
DMFT provides a recipe to calculate these two-particle quantities~\cite{Khurana90,Georges96} and that recipe starts with the two-particle properties of the single-site impurity model. 

There is an additional reason why the two-particle properties of the impurity are of interest. In finite-dimensional systems, DMFT is an approximation since it only captures local correlations. Extending DMFT to add non-local correlations is currently an area of active interest~\cite{Maier05,Rohringer18}. Many of these approaches use the two-particle properties of the impurity model as a starting point. This has motivated a systematic investigation of issues like their frequency structure, asymptotics and symmetries~\cite{Kunes11,Rohringer12,Schafer13,Wentzell16,Tagliavini18,Reckling18}.

So far, these investigations have mainly been focussed on the fermion-fermion vertex $\gamma$. The fermion-boson vertex $\lambda$, which is effectively obtained by tracing out one degree of freedom in the fermion-fermion vertex, has received less attention even though it is simpler. Studying this simplified object is relevant, since it describes the system's response to external fields~\cite{Migdal67}. The vertex also  describes the coupling between the fermions and (effective) bosons such as plasmons, magnons, orbitons or Cooper pairs. In this way, the vertex appears in Hedin's equations~\cite{Hedin65,Aryasetiawan98} and in functional renormalization group descriptions~\cite{Bartosch09,Friederich10,Metzner12}.

Furthermore, the fermion-boson vertex appears in several diagrammatic extensions of DMFT, such as dual boson~\cite{Rubtsov12} and TRILEX~\cite{Ayral15}. In fact, the simplicity of the fermion-boson vertex was one of the motivations for the TRILEX~\cite{Ayral15,Ayral16} method: approximating the fermion-boson vertex of the lattice model by the vertex of a self-consistent impurity model.
In the dual boson approach, the fermion-boson vertex determines the leading order diagrammatic contributions. It has been shown~\cite{vanLoon14} that the fermion-boson vertex is sufficient to capture the leading non-local correlation effects in the charge-density wave transition. The fermion-boson vertex is also suffices to obtain the correct high-frequency asymptote of the momentum-resolved susceptibility in dual boson~\cite{Stepanov16,Krien17}. 
Within DMFT, the fermion-boson vertex can be used to efficiently evaluate the spatial susceptibility~\cite{Pruschke96}.
The advantage of the fermion-boson vertex is even clearer with an eye on multi-orbital systems, since the reduced object carries only two fermionic orbital labels and one bosonic channel index. This makes application of the fermion-boson vertex more computationally feasible, especially if the relevant channels (e.g., density, magnetic or a specific orbital combination) are known a priori. 

In this work, we systematically study the fermion-boson vertex in DMFT. We study its physical content, analytical properties, perturbative expansion at small interaction strength, Ward identity, asymptotics and its behaviour near the metal-insulator transition. Compared to previously published results for the charge channel~\cite{Hafermann14,vanLoon14}, we add the magnetic channel and discuss similarities and differences between the two. We also extend results obtained for the fermion-fermion vertex to the fermion-boson vertex.

\section{Physical content of the fermion-boson correlation function}
\label{sec:physicalcontent}

Two-particle functions play an important role in our theoretical understanding of interacting systems. They have two physical interpretations. On the one hand, they can be interpreted as describing the \emph{correlation} in a system. On the other hand, the susceptibility describes the (linear) \emph{response} of the system to external stimuli. These two meanings are linked together by the Kubo formula~\cite{Kubo57}, mathematically both correlation and response are obtained as second derivatives of the free energy. 

An illustrative example of a two-particle function is the magnetic susceptibility, 
\begin{align*}
\frac{d\av{S^z_1}}{dh_2} = \av{S^z_1 S^z_2} - \av{S^z_1}\av{S^z_2}, 
\end{align*}
which involves two space-time coordinates, which corresponds to one frequency and momentum in a translationally invariant system. As a response function, this tells us how the magnetization at coordinate 1 changes when we apply a magnetic field at coordinate 2. The equality of correlation and response function follows from the fact that the magnetic field $h$ is conjugate to $S^z$ in the Hamiltonian, i.e., $H=H_0 -h S^z$.

Since we are considering systems of fermions, the (bosonic) operator $S^z$ is an operator consisting of pairs of fermionic operators. The generalized concept of a two-particle susceptibility thus consists of a four-fermion correlation function, with four space-time coordinates or three frequencies and momenta,
\begin{align*}
 \av{c^{\phantom{*}}_1 c^*_2 c^{\phantom{*}}_3 c^*_4}.
\end{align*}
As a response function, this corresponds to the change in $\av{c_1 c^*_2}$ due to a field that removes a fermion at coordinate 3 and puts it back at coordinate 4. From a concrete physical point of view, this response function is hard to visualize.  

The subject of this work considers an intermediate option, where only one of the bosonic operators has been replaced by the underlying fermions~\cite{Migdal67} (the minus sign here is a question of convention),
\begin{align}
 -\av{c^{\phantom{*}}_1 c^*_2 S^z_3}+\av{c^{\phantom{*}}_1 c^*_2}\av{S^z_3}  =& \frac{dG_{12}}{dh_3},\label{eq:L:abstract}\\
 G_{12} =& -\av{c^{\phantom{*}}_1 c^*_2}. 
\end{align}
Just like the magnetic susceptibility, this quantity describes the response to a magnetic field. The difference is that the observable under investigation is not simply a number, it is an object which has internal coordinates. Transforming to frequency space, the Green's function describes the spectral function of the system and the response function tells us how the spectral function changes when an external magnetic stimulus is applied~\cite{Migdal67}. This stimulus can be a static, homogeneous field, but it can also be a periodic field or a magnetic adatom on top of a two-dimensional surface. The fermion-boson vertex plays a fundamental role in Fermi liquid theory~\cite{Migdal67}.

To be more concrete, let us start by considering a system of non-interacting paramagnetic fermions with spin $\up$ and $\dn$. If we apply a homogeneous, time-independent magnetic field $h$, what effectively happens is that all the energy levels of the $\up$ and $\dn$ spins move in opposite directions and the paramagnetism is broken. After integrating over the energy of the fermions, the usual magnetic susceptibility $dm/dh$ is found, which gives the total change in magnetization.

The situation at finite frequency is slightly different. A homogeneous finite frequency field does not change the net magnetization at any time since both the number of $\up$ and $\dn$ fermions are conserved quantities. Even though it does not change the number of particles, the field does have an effect on the spectrum. A fermionic state $a$ with energy $E_a<0$ would normally show up in the photoemission spectrum exactly at energy $E_a$. In the presence of a finite frequency field, however, it is possible for the fermion to absorb (or emit) a quantum of energy $\omega$ from the field $h_\omega$, so that it now shows up at energy $E_a\pm \omega$ in the photoemission spectrum. We are restricting our analysis to linear response, which means that only a single quantum of energy is emitted or absorbed. The overall magnitude of these shifted features in the photoemission spectrum is proportional to $h$. 

To be more precise, we can use the fermion-boson vertex of the non-interacting system to see what happens.
Starting from the paramagnetic solution, the linear response is (this type of calculation will be done in more detail in the remainder of the manuscript, here it serves to illustrate the general procedure):
\begin{align}
 \frac{dG_{\up,E,E+\omega} }{dh_\omega} =& -\av{c^{\phantom{*}}_{\up,E} c^*_{\up,E+\omega} S^z_{\omega}}+\av{c^{\phantom{*}}_{\up,E} c^*_{\up,E+\omega}}\av{S^z_{\omega}} \notag \\
 =& -G^0_{\up,E}G^0_{\up,E+\omega}, \notag
\end{align}
where the second line follows from Wick's theorem after using $S^z=c^*_\up c^{\phantom{*}}_\up - c^*_\dn c^{\phantom{*}}_\dn$. A subtlety is that the Green's function on the left-hand side now needs two frequency arguments, the energy of a single fermion is not conserved since it can exchange energy with the external field. In terms of times instead of energies, if an external field is applied at $t_0$, time translation symmetry is broken and the Green's function depends not just on $t_2-t_1$ but on both times separately. On the right-hand side, everything is written in terms of the system without external field, where a single frequency argument is sufficient, since the system without external field has time translation symmetry.
For the density of states $A_E=-\frac{1}{\pi}\Im G_E$, we thus find
\begin{align}
 \frac{dA_{\up,E,E+\omega} }{dh_\omega} = \Re G^0_{\up,E+\omega} \cdot A^0_{\up,E} + \Re G^0_{\up,E} \cdot A^0_{\up,E+\omega}.
\end{align}
The change in the photoemission spectrum has two contributions, the first from particles that were originally at $E$ and have absorbed energy $\omega$, the other from particles that were at $E+\omega$ and emit energy $\omega$.  

This was the situation for a non-interacting system, where physical intuition already explains what will happen. In an interacting system, the situation is more subtle. First of all, the interaction changes the properties of the fermion quasiparticles which leads to energy shifts and finite lifetimes. Secondly, the response to the external field will also change. The particle that absorbs a quantum of energy from the field will be interacting with all other fermions. The fermion-boson vertex quantifies these interaction effects. 

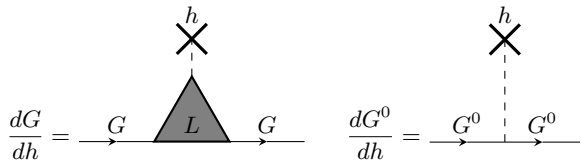
\begin{figure}
$\displaystyle \frac{dG}{dh}= $
\begin{tikzpicture}


\draw [->] (0,0) -- (0.5,0) ;
\draw[-] (0.5,0) -- (1,0);
\draw [thick,fill=black!50] (1,0) -- ++(1,0) -- coordinate[pos=1] (top) ++(120:1) -- cycle ;
\draw [->] (2,0) -- (2.5,0) ;
\draw [-] (2.5,0) -- (3,0) ;

\draw[dashed] (top) -- coordinate[pos=1] (cross) ++(0,0.5) ;

\draw[very thick] (cross) -- ++( 0.2, 0.2) ;
\draw[very thick] (cross) -- ++( 0.2,-0.2) ;
\draw[very thick] (cross) -- ++(-0.2, 0.2) ;
\draw[very thick] (cross) -- ++(-0.2,-0.2) ;

\node[above] at (0.5,0) {$G$} ;
\node[above] at (2.5,0) {$G$} ;
\node[above] at (1.5,0) {$L$} ;
\node[above] at ($(cross)+(0,0.1)$) {$h$} ;

\end{tikzpicture}%
\,\,\, \,\,\,
$\displaystyle \frac{dG^0}{dh}= $
\begin{tikzpicture}%
\draw [->] (0,0) -- (0.5,0) ;
\draw[-] (0.5,0) -- (1,0);
\draw [->] (1,0) -- (1.5,0) ;
\draw [-] (1.5,0) -- (2,0) ;

\draw[dashed] (1.0,0) -- coordinate[pos=1] (cross) ++(0,.5+0.866) ;

\draw[very thick] (cross) -- ++( 0.2, 0.2) ;
\draw[very thick] (cross) -- ++( 0.2,-0.2) ;
\draw[very thick] (cross) -- ++(-0.2, 0.2) ;
\draw[very thick] (cross) -- ++(-0.2,-0.2) ;

\node[above] at (0.5,0) {$G^0$} ;
\node[above] at (1.5,0) {$G^0$} ;
\node[above] at ($(cross)+(0,0.1)$) {$h$} ;

\end{tikzpicture}
\caption{The fermion-boson vertex~\cite{Migdal67} $L$ describes the linear response to an external field $h$ in diagrammatic notation. In a non-interacting system, it is equal to unity.}
\label{fig:migdal}
\end{figure}

Having seen the non-interacting result, a product of two Green's functions, it makes sense to write 
\begin{align}
 \frac{dG_{E,E+\omega} }{dh_\omega} = G_E L_{E,\omega} G_{E+\omega} \label{eq:L:abstract2}
\end{align}
which serves as the definition of $L$, c.f., Eq.~\ref{eq:L:abstract}. The non-interacting system then corresponds simply to $L=-1$ (the minus sign is a question of convention). The interacting and non-interacting response to the external field can be illustrated diagrammatically as in Fig.~\ref{fig:migdal} (see Chapter 2 of Ref.~\onlinecite{Migdal67} for a pedagogical introduction). 

\section{Model and Method}

We now move towards a concrete model and method where the fermion-boson vertex can be determined numerically.
Dynamical Mean-Field Theory (DMFT) provides a description of local correlations in interacting fermion systems. This is done by mapping the interacting lattice model to an effective single-site impurity model. We determine the fermion-boson vertex of this impurity model. The parameters of the effective impurity model are determined in a self-consistent fashion, in such a way that the impurity model mimics the original lattice model.

To be more concrete, we consider the Hubbard model on the triangular lattice. This model is given by the Hamiltonian
\begin{align}
 H = -\sum_{ij,\sigma} t_{ij} c^\dagger_{i\sigma} c^{\phantom{\dagger}}_{j\sigma} + U \sum_{i} n_{i\up} n_{i\dn},  
\end{align} 
where $c^\dagger_{i\sigma}$ is the creation operator for a fermion on site $i$ with spin $\sigma =\up,\dn$. $n_{i\sigma}= c^\dagger_{i\sigma} c^{\phantom{\dagger}}_{i\sigma}$ counts the number of electrons with spin $\sigma$ on site $i$, due to the Pauli principle  $n_{i\sigma}$ is either 0 or 1. For the hopping $t_{ij}$, we only consider hopping between nearest-neighbors in the triangular lattice and in the following we use the nearest-neighbor matrix element $t=1$ as the unit of energy. Any on-site energy $t_{ii}$ is incorporated in the chemical potential. The interaction $U$ between fermions of opposite spin happens when they simultaneously occupy the same site. In electronic systems, typically $U>0$ since the Coulomb interaction between electrons is repulsive. In other fermionic systems, $U<0$ is a possibility and our analysis holds for both situations. We study this Hubbard model as a function of temperature $1/\beta$ and chemical potential $\mu$. 

The DMFT impurity model is given in the action formulation as
\begin{align}
S_{\text{imp}} = - \frac{1}{\beta} \sum_{\nu\sigma} c^{*}_{\nu\sigma} (i\nu+\mu-\Delta_\nu) c^{\phantom{*}}_{\nu\sigma} + U n_\up n_\dn, \label{eq:Simp}
\end{align}
where $\nu$ denotes a fermionic Matsubara frequency.
Here, the hybridization $\Delta$ is chosen in a self-consistent fashion, using the condition that the local Green's function of the impurity model and the associated lattice model should be equal:
\begin{align}
 g_{\nu}^{\sigma} = \left(G^{\text{DMFT}}_{\text{local}}\right)_\nu^{\sigma}
\end{align}
where $g_{\nu}^\sigma = -\av{c^{\phantom{*}}_{\nu\sigma} c^*_{\nu\sigma} }$ is the Fourier transform of the imaginary time Green's function of the impurity model and $\av{\cdot}$ denotes the expectation value of an observable within the impurity model, Eq.~\eqref{eq:Simp}. 
The DMFT Green's function of the lattice model is given by 
$\left(G^{\text{DMFT}}_{\text{local}}\right)_\nu^{\sigma} = \sum_{\kv} (g^\sigma_\nu+\Delta_\nu-t_\kv)^{-1}$, where $\sum_\kv$ is the average over momentum $\kv$ and $t_\kv$ is the Fourier transform of $t_{ij}$.
Note that $t_\kv$ is the only place where the lattice structure enters the DMFT formalism.
A more detailed explanation of DMFT can be found in the review~\cite{Georges96}.

As is usual for mean-field theories, information about the actual lattice structure enters the calculation only in a rather coarse way. For example, for Weiss' mean-field theory for magnetism, only the coordination number enters into the calculations. In the case of DMFT, it is the local density of states (DOS) of the non-interacting model that carries the information about the lattice into the computations. In that sense, our results here should mostly be considered to describe the rather generic phenomena described by the DMFT approximation rather than specific features of the triangular lattice. In particular, DMFT does not provide an accurate description of specifically low dimensional phenomena like magnetic ordering.

The triangular lattice was chosen since it does not have special additional symmetries, like the particle-hole symmetry of bipartite lattices. There is no vanishing of the density of states, as in the honeycomb lattice, and the Van Hove singularity is well separated from half-filling where the Mott-Hubbard transition occurs.

Coming back to the impurity problem, we use a continuous-time quantum Monte Carlo (CTQMC) solver~\cite{Rubtsov05,Werner06,Gull11} based on the hybridization expansion. This solver~\cite{Hafermann13} uses improved estimators~\cite{Hafermann14} for the fermion-boson vertex: the Monte Carlo routine measures higher-order correlation correlation functions that are related to the fermion-boson vertex via the equation of motion.

\section{fermion-boson vertex of the impurity model}
\label{sec:definition}

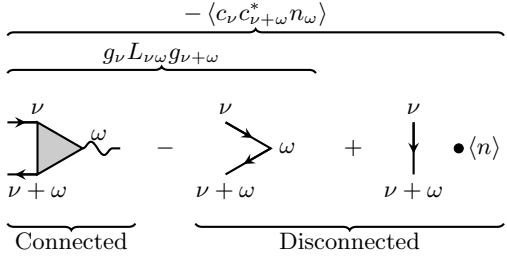
\begin{figure} 
\begin{tikzpicture}

\coordinate (b1) at (  0:0.4cm) ;
\coordinate (b2) at (120:0.4cm) ;
\coordinate (b3) at (240:0.4cm) ;

\draw[thick,fill=black!20] (b1) -- (b2) -- (b3) -- cycle ;

\draw[thick,-]  (b2) -- coordinate[pos=0.7] (top)  ++(-0.4,0) ;
\draw[thick,-<]  (b2) -- ++(-0.3,0) ;

\draw[thick,-]  (b3) -- coordinate[pos=0.7] (bottom) ++(-0.4,0) ;
\draw[thick,->]  (b3) -- ++(-0.3,0) ;

\draw[decorate,decoration={snake},thick] (b1) -- coordinate[pos=0.5] (right) ++(0.5,0);

\node[above] at (b2) {$\nu$};
\node[below] at (b3) {$\nu+\omega$};
\node[above right] at (b1) {$\omega$};

\begin{scope}[shift={(1.5,0)}]
 \node at (0,0) {$-$} ;
\end{scope}

\begin{scope}[shift={(2.5,0)}]
\coordinate (c1) at (  0:0.4cm) ;
\coordinate (c2) at (120:0.4cm) ;
\coordinate (c3) at (240:0.4cm) ;

\draw[thick] (c2) -- (c1) -- (c3) ;
\draw[thick,->] (c2) -- ($(c1)!0.4!(c2)$) ;
\draw[thick,->] (c1) -- ($(c1)!0.6!(c3)$) ;

\node[above] at (c2) {$\nu$};
\node[below] at (c3) {$\nu+\omega$};
\node[right] at (c1) {$\omega$};

\end{scope}

\begin{scope}[shift={(4,0)}]
 \node at (0,0) {$+$} ;
\end{scope}

\begin{scope}[shift={(5,0)}]
\coordinate (d1) at (  0:0.4cm) ;
\coordinate (d2) at (120:0.4cm) ;
\coordinate (d3) at (240:0.4cm) ;

\draw[thick] (d2) -- (d3) ;
\draw[thick,->] (d2) -- ($(d2)!0.6!(d3)$) ;

\filldraw (d1) circle (0.06cm);

\node[above] at (d2) {$\nu$};
\node[below] at (d3) {$\nu+\omega$};
\node[right] at (d1) {$\av{n}$};

\end{scope}

\draw [thick, decorate, decoration={brace}] (-0.6,1.5) -- node[above] {$-\av{c_\nu c^*_{\nu+\omega} n_\omega}$} (6,1.5) ;

\draw [thick, decorate, decoration={brace}] (-0.6,1) -- node[above] {$g_\nu L_{\nu\omega} g_{\nu+\omega}$} (3.5,1) ;

\draw [thick, decorate, decoration={brace,mirror}] (1.9,-1) -- node[below] {Disconnected} (6,-1) ;

\draw [thick, decorate, decoration={brace,mirror}] (-0.6,-1) -- node[below] {Connected} (1.1,-1) ;

\end{tikzpicture}
\caption{Diagrammatic structure of the fermion-boson vertex $L$, c.f. Ref.~\onlinecite{vanLoon14}. Straight lines with arrows denote the single-particle Green's function $g$, the wavy line denotes the coupling to the external bosonic field. Note that the vertex here is rotated 90 degrees with respect to Fig.~\ref{fig:migdal}, the boson is on the right-hand side instead of at the top.}
\label{fig:diagrammaticcontent}
\end{figure}

With that, we come to the main subject of this work, the fermion-boson vertex of the impurity model. Several definitions are possible, especially regarding the normalization. The physical content is independent of these definitions. For our purposes here, the most convenient definition is
\begin{align}
    L^{\sigma'\sigma}_{\nu\omega} =& \frac{- \av{c^{\phantom{*}}_{\nu\sigma} c^*_{\nu+\omega,\sigma} n^{\phantom{*}}_{\sigma',\omega} } - \beta g_{\nu} \av{n_{\sigma'}} \delta_\omega }{g_\nu g_{\nu+\omega} } \label{eq:lambda:spinresolved}.
\end{align}
Here, $L^{\sigma'\sigma}$ is the coupling between a $\sigma$ electron and an $\sigma'$ boson, where $n_{\sigma',\omega}=\sum_{\nu'} c^\dagger_{\sigma',\nu'} c^{\phantom{\dagger}}_{\sigma',\nu'+\omega}$. The fermionic and bosonic Matsubara frequencies $\nu$ and $\omega$ should be understood as the Fourier transform of the imaginary time correlation functions. In the following we assume paramagnetism so that it is sufficient to consider $L^{\up\up}$ and $L^{\up\dn}$. The definition of $L$ is illustrated in Fig.~\ref{fig:diagrammaticcontent}.

The original dual boson works~\cite{Rubtsov12,vanLoon14} use a slightly different definition. There, the fermion-boson vertex is given by
\begin{align}
 \lambda^{\rho,\sigma}_{\nu\omega} = \frac{-\av{c^{\phantom{*}}_{\nu\sigma} c_{\nu+\omega,\sigma}^{*} \rho_\omega} - \beta g^\sigma_\nu \av{\rho}\delta_\omega }{g_{\nu}^{\sigma}g_{\nu+\omega}^{\sigma}\chi^\rho_{\omega} }, \label{eq:lambda:def}
\end{align}
where $\rho$ is a Hermitian bosonic variable, we restrict our attention to $\rho=n=n_\up+n_\dn$ or $\rho=S^z=n_\up - n_\dn$, and
\begin{align}
 \chi^\rho_\omega=& -\av{\rho_\omega \rho_{-\omega}} + \av{\rho}\av{\rho}\delta_\omega,
\end{align}
is the corresponding susceptibility of the impurity model. This definition differs in two ways from Eq.~\eqref{eq:lambda:spinresolved}. First, there is a normalization by $\chi^{-1}$. Secondly, $\lambda$ is defined in the charge and $S^z$ channel whereas $L$ is defined per spin component. The two are related as
\begin{align}
    L^{\up\up}_{\nu\omega} =& \frac{1}{2} \left( \chi^n_\omega \lambda^{n,\up}_{\nu\omega} + \chi^z_\omega \lambda^{z,\up}_{\nu\omega}\right), \notag \\
    L^{\dn\up}_{\nu\omega} =& \frac{1}{2} \left( \chi^n_\omega \lambda^{n,\up}_{\nu\omega} - \chi^z_\omega \lambda^{z,\up}_{\nu\omega}\right) \notag
\end{align}
Essentially, the spin-based notation is the appropriate basis for the non-interacting system where both spin sectors are decoupled. On the other hand, the charge and magnetic channel are the best basis for the Hubbard interaction, as can be seen in the RPA expression for the susceptibility and in the observation that the metal-insulator transition corresponds to a gap opening in the charge sector and not in the spin sector.

\begin{figure}
\begin{tikzpicture}
\begin{scope}[xscale=-1.3,yscale=1.3]
\draw[thick,->] (-.1,-0.15) -- (0.25,-0.15) ;
\draw[thick,->] (0.55,-0.15) -- (1,-0.55) ;
\draw[thick,->] (1,0.55) -- (0.55,0.15) ;
\draw[thick,->] (0.25,0.15) -- (-.1,0.15) ;

\filldraw[thick,black!25] (0.25,-0.15) -- (0.55,-0.15) -- (0.55,0.15) -- (0.25,0.15);
\node[] at (0.4,0) {$\gamma$}; 

\draw [thick, decorate, decoration={brace,mirror}] (-0.15,-0.25) -- (-0.15,0.25) ;

\node[left] at (1,0.55) {$\sigma,\,\nu$} ;
\node[left] at (1,-0.55) {$\sigma,\,\nu+\omega$} ;
\node[anchor=west,text width=1.3cm,align=left] at (-0.2,0) {$\sigma',\,\nu'$\linebreak$\sigma',\,\nu'\!+\!\omega$};
\end{scope}
\end{tikzpicture}
\caption{The connected contribution to the fermion-boson vertex $L$, where $\gamma$ is the fermion-fermion vertex. The two fermionic propagators on the right-hand side together comprise the boson line of Fig.~\ref{fig:diagrammaticcontent}, as indicated by the brace.}
\label{fig:Lgamma} 
\end{figure}
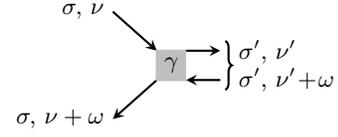

The fermion-boson vertex is related to the fermion-fermion vertex $\gamma$,
\begin{align}
 \gamma^{\sigma\sigma'}_{\nu\nu'\omega} =&
 \frac{
 g^{(4)\sigma\sigma'}_{\nu\nu'\omega}
 + \beta g_{\nu\sigma} g_{\nu+\omega\sigma} \delta_{\nu\nu'}\delta_{\sigma\sigma'}
 -\beta g_{\nu\sigma} g_{\nu'\sigma' \delta_\omega} 
 }{
 g_{\nu\sigma}g_{\nu+\omega\sigma} g_{\nu'\sigma'}g_{\nu'+\omega\sigma'}
 } \notag \\
 g^{(4)\sigma\sigma'}_{\nu\nu'\omega} =& +\av{
 c^{\phantom{*}}_{\nu\sigma} 
 c^{*}_{\nu+\omega\sigma}
 c^{\phantom{*}}_{\nu'\sigma'}
 c^{*}_{\nu'+\omega\sigma'}
 }\label{eq:def:gamma}.
\end{align}
The fermion-boson vertex can be obtained from the fermion-fermion vertex by ``tying together'' two fermion lines of $\gamma$~\cite{Migdal67,Rubtsov12}. This corresponds to summing over $\nu'$ in Eq.~\eqref{eq:def:gamma}. In a previous work~\cite{vanLoon14}, we used only the charge channel, here we also give the explicit variant for the spin channel,
\begin{align}
\lambda_{\nu\omega}^{\rho,\sigma} = \left(\chi^\rho_{\omega}\right)^{-1} \left(\frac{1}{\beta}\sum_{\sigma'\nu'}
\left(\pm 1\right)^{\rho,\sigma,\sigma'}
\gamma^{\sigma\sigma'}_{\nu\nu'\omega}g_{\nu'}^{\sigma'}g_{\nu'+\omega}^{\sigma'} -1\right).
\end{align}
Here, $(\pm 1)^{\rho,\sigma,\sigma'}$ is equal to 1 for $\rho=n$, for $\rho=S^z$ it is equal to $+1$ if $\sigma=\sigma'$ and $-1$ otherwise.
The spin structure is clearer when written in the notation of Eq.~\eqref{eq:lambda:spinresolved},\footnote{Note that in our notation, in $L^{\sigma'\sigma}_{\nu\omega}$ the fermion with spin $\sigma$ has frequency $\nu$ and the fermion with spin $\sigma'$ has the frequency $\nu'$ which is summed over. Compared to the definition for $\gamma$ the spin labels are interchanged, as is visible in $L^{\dn\up}$.}
\begin{align}
L_{\nu\omega}^{\up\up} =& {-1}+\frac{1}{\beta}\sum_{\nu'}
\gamma^{\up\up}_{\nu\nu'\omega}g_{\nu'}g_{\nu'+\omega} \notag \\
L_{\nu\omega}^{\dn\up} =& \phantom{-1}+\frac{1}{\beta}\sum_{\nu'}
\gamma^{\up\dn}_{\nu\nu'\omega}g_{\nu'}g_{\nu'+\omega}. \label{eq:L:gamma}
\end{align}
These formulas show that even in a non-interacting system, i.e., $U=0$ and $\gamma=0$, the fermion-boson vertex has a finite value of $-1$. This finite value comes from the equal-spin channel. 
In terms of response functions, this corresponds exactly to the response of a non-interacting system, as discussed in Sec.~\ref{sec:physicalcontent}.

The relation between the fermion-fermion and fermion-boson vertex follows from the fact that the boson in question actually consists of two fermions. Diagrammatically, the wavy line in Fig.~\ref{fig:diagrammaticcontent} is replacement by two fermionic lines. The difference is that one of the fermionic frequencies, $\nu'$ is summed over, these two fermions are tied together into a bosonic operator.\footnote{In $\tau$-space, these lines have the same value of $\tau$.}  There are two disconnected diagrams in Fig.~\ref{fig:diagrammaticcontent}.
The disconnected diagram on the right is removed by the term proportional to $\delta_\omega$ in Eq.~\eqref{eq:lambda:def}. In the diagram on the left, spin and energy conservation require $\sigma'=\sigma$ and $\nu'=\nu$. In that case, it is equal to $g_{\nu}g_{\nu+\omega}$, which cancels with the denominator of Eq.~\eqref{eq:lambda:def} and gives the trivial non-interacting contribution $-1$ to the vertex. The non-trivial contribution is obtained by inserting the exact two-particle vertex $\gamma$ into the diagram, as shown in Fig.~\ref{fig:Lgamma}.

\section{Perturbative analysis}
\label{sec:diagrams}

\begin{table*}
\begin{tabular}{c | c | c | c | c}
$ L^{\dn\up}_{\nu\omega}|^\text{(a)}$ & $L^{\dn\up}_{\nu\omega}|^\text{(b)}$ & $L^{\dn\up}_{\nu\omega}|^\text{(b')}$ & $L^{\up\up}_{\nu\omega}|^\text{(c)} $ & $L^{\up\up}_{\nu\omega}|^\text{(d)}$ \\
\hline
\hline
&&&&\\
$\displaystyle -U B_\omega$ &
$\displaystyle +\frac{U^2}{\beta} \sum_{\nu'} g_{\nu'}g_{\nu'+\omega} B_{\nu'-\nu} $ &
$\displaystyle +\frac{U^2}{\beta} \sum_{\nu'} g_{\nu'}g_{\nu'+\omega} B^{pp}_{\nu+\nu'+\omega} $ &
$\displaystyle -\frac{U^2}{\beta} \sum_{\nu'}
g_{\nu'}g_{\nu'+\omega} B_\omega $ &
$\displaystyle +\frac{U^2}{\beta}\sum_{\nu'}
g_{\nu'}g_{\nu'+\omega} B_{\nu'-\nu} $ \\
&&&&\\
\begin{tikzpicture}
\begin{scope}[xscale=-1]
\draw[thick,->,fmRed] (0,-0.3) -- (0.3,0) ;
\draw[thick,<-,fmRed] (0,+0.3) -- (0.3,0) ;
\draw[thick,->,fmBlue] (0.3,0) -- (0.6,-0.3) ;
\draw[thick,<-,fmBlue] (0.3,0) -- (0.6,0.3) ;

\draw [thick, decorate, decoration={brace,mirror}] (-0.15,-0.3) -- (-0.15,0.3) ;

\end{scope}
\end{tikzpicture}&
\begin{tikzpicture}
\begin{scope}[xscale=-1]
\draw[thick,->,fmRed] (0,-0.6) -- (0.3,-0.3) ;
\draw[thick,<-,fmRed] (0,+0.3) -- (0.3,0) ;
\draw[thick,->,fmBlue] (0.3,-0.3) -- (0.6,-0.6) ;
\draw[thick,<-,fmBlue] (0.3,0) -- (0.6,0.3) ;
\draw[thick,->,fmRed,out=90+45,in=-90-45,looseness=2] (0.3,-0.3) to (0.3,0) ;
\draw[thick,<-,fmBlue,out=45,in=-45,looseness=2] (0.3,-0.3) to (0.3,0) ;

\draw [thick, decorate, decoration={brace,mirror}] (-0.15,-0.6) -- (-0.15,0.3) ;

\end{scope}
\end{tikzpicture}&
\begin{tikzpicture}
\begin{scope}[xscale=-1]
\draw[thick,->,fmRed] (0,-0.6) -- (0.3,-0.3) ;
\draw[thick,<-,fmRed] (0,+0.3) -- (0.3,0) ;
\draw[thick,->,fmBlue,out=45,in=180] (0.3,0) to (0.9,-0.6) ;
\draw[thick,<-,fmBlue,out=-45,in=180] (0.3,-0.3) to (0.9,+0.3) ;
\draw[thick,->,fmRed,out=90+45,in=-90-45,looseness=2] (0.3,-0.3) to (0.3,0) ;
\draw[thick,->,fmBlue,out=45,in=-45,looseness=2] (0.3,-0.3) to (0.3,0) ;

\draw [thick, decorate, decoration={brace,mirror}] (-0.15,-0.6) -- (-0.15,0.3) ;
\end{scope}
\end{tikzpicture}&
\begin{tikzpicture}
\begin{scope}[xscale=-1]
\draw[thick,->,fmBlue] (0,-0.3) -- (0.3,0.0) ;
\draw[thick,<-,fmBlue] (0,+0.3) -- (0.3,0) ;
\draw[thick,->,fmBlue] (0.6,0) -- (0.9,-0.3) ;
\draw[thick,<-,fmBlue] (0.6,0) -- (0.9,0.3) ;
\draw[thick,->,fmRed,out=-45,in=-90-45,looseness=2] (0.3,0.) to (0.6,0) ;
\draw[thick,<-,fmRed,out=45,in=90+45,looseness=2] (0.3,0.) to (0.6,0) ;

\draw [thick, decorate, decoration={brace,mirror}] (-0.15,-0.3) -- (-0.15,0.3) ;
\end{scope}
\end{tikzpicture}&
\begin{tikzpicture}
\begin{scope}[xscale=-1]
\draw[thick,->,fmBlue] (0,-0.6) -- (0.3,-0.3) ;
\draw[thick,<-,fmBlue] (0,+0.3) -- (0.3,0) ;
\draw[thick,->,fmBlue] (0.3,-0.3) -- (0.6,-0.6) ;
\draw[thick,<-,fmBlue] (0.3,0) -- (0.6,0.3) ;
\draw[thick,->,fmRed,out=90+45,in=-90-45,looseness=2] (0.3,-0.3) to (0.3,0) ;
\draw[thick,<-,fmRed,out=45,in=-45,looseness=2] (0.3,-0.3) to (0.3,0) ;

\draw [thick, decorate, decoration={brace,mirror}] (-0.15,-0.6) -- (-0.15,0.3) ;
\end{scope}
\end{tikzpicture}\\
s-channel&t-channel&u-channel&s-channel&t-channel
\end{tabular}
\caption{Diagrams contributing to the fermion-boson vertex $L$ up to order $U^2$. Blue lines are $\up$ fermions, red lines are $\dn$ fermions. The two outgoing lines on the right-hand side of the diagram together form the bosonic variable in the vertex, as indicated by the brace. We use labels (a)-(d) to refer to these diagrams in the text. Note that only in the expression for diagram (c) the sum over $\nu'$ of the two Green's functions could be replaced by the bubble $B_\omega$, this was not done here to keep the expressions similar. The channels are classified according to the Mandelstam variables~\cite{Mandelstam58} s, t and u. The external frequency and spin labels are the same as in Fig.~\ref{fig:Lgamma}.
}
\label{table:diagrams}
\end{table*}

The relation between $L$ and $\gamma$, Eq.~\eqref{eq:L:gamma}, allows us to transfer the diagrammatic analysis~\cite{Rohringer12,HummelThesis} of $\gamma$ in terms of $U$ into an expansion for $\lambda$ and $L$. In this weak-coupling expansion, we do use the exact single-particle Green's function $g$ and susceptibility $\chi$. The idea here is that $g$ and $\chi$ are much easier to determine than $\gamma$, $\lambda$ and $L$, since the former depend only on a single frequency. Doing the expansion in terms of the interacting $g$ means that we do not have to take into account self-energy insertion diagrams.
Using $L$, which has explicit spin labels, is beneficial since the bare interaction $U$ only couples electrons with opposite spin. This interaction happens instantaneously.

The Feynman rules used for drawing these diagrams are given in Appendix~\ref{app:feynmanrules}. Simply put, we use blue and red lines for fermionic propagators $g$ with spin up and down respectively. The 4-fermion Hubbard interaction $U$ has one ingoing and one outgoing line of both spin flavors. The results of the diagrammatic expansion that follows are summarized in Table~\ref{table:diagrams}, where the various diagrams are denoted by letters and will be discussed in the following.

\subsubsection{$L^{\up\dn}$}
Starting with the unequal spin vertex, we obtain one diagram that features a single Hubbard interaction,
\begin{align}
 L^{\dn\up}_{\nu\omega}|^{\text{(a)}} =& -\frac{U}{\beta} \sum_{\nu'} g_{\nu'}g_{\nu'+\omega} \\
 =& -U B_{\omega}, 
\end{align}
where $B$ is a shorthand notation for the bubble of two Green's function, see Appendix~\ref{app:bubble}. 
This result is notably independent of $\nu$.  This expression can also be obtained by inserting the lowest-order approximation for the vertex,
$ \gamma^{\dn\up}_{\nu\nu'\omega} \approx -U $
, into Eq.~\eqref{eq:L:gamma}.

The two Green's functions in this expression appear as a bubble, in other words as $\chi^{0}$. We can account for higher orders in $U$ by replacing $\chi^{0}$ by the appropriate susceptibility $\chi$, taking into account the proper spin labels. A practical advantage is that $\chi_\omega$ can be measured directly in the impurity solver, whereas the bubble $B_\omega$ has to be obtained as a sum over fermionic frequency $\nu'$ with a finite frequency cutoff.
\begin{align}
 L^{\dn\up}_{\nu\omega}|^{\text{(a)}} = & -U \chi_\omega^{\dn\dn} \notag \\
 = & -U (\chi_\omega^c+\chi_\omega^z)/4. \label{eq:lambda:unequalU}
\end{align}
Here $\chi^{\dn\dn} = -\av{n_\dn n_\dn}+\av{n_\dn}\av{n_\dn}$ is the dressed propagator of a particle-hole pair of $\dn$ electrons, diagrammatically \linebreak
\begin{tikzpicture}
\begin{scope}[xscale=-1]
\draw[thick,->,fmRed] (0,-0.3) -- (0.3,0) ;
\draw[thick,<-,fmRed] (0,+0.3) -- (0.3,0) ;
\draw[thick,->,fmBlue] (0.3,0) -- (0.6,-0.3) ;
\draw[thick,<-,fmBlue] (0.3,0) -- (0.6,0.3) ;
\end{scope}
\end{tikzpicture}
 turns into %
\begin{tikzpicture}
\begin{scope}[xscale=-1]
\draw[thick,fmRed] (0,-0.3) -- (0.3,0) ;
\draw[thick,fmRed] (0,+0.3) -- (0.3,0) ;
\draw[thick,->,fmBlue] (0.3,0) -- (0.6,-0.3) ;
\draw[thick,<-,fmBlue] (0.3,0) -- (0.6,0.3) ;
\draw[thick,->,fmRed] (-0.3,-0.3) -- (0,-0.3) ;
\draw[thick,<-,fmRed] (-0.3,0.3) -- (0,0.3) ;

\begin{scope}[even odd rule]
 \clip (0.15,0.3) -- (0.15,-0.3) -- (0,-0.3) -- (0,0.3) -- cycle;
 \fill[fmRed] (0.3,0) -- (0,-0.3) -- (0,0.3) -- cycle ; 
\end{scope}

\end{scope}
\end{tikzpicture}
.
Here, we should remember that the blue lines at the end, $g_{\nu\up}$, $g_{\nu+\omega\up}$, are divided out (amputated) in the expression for $L$, so that there is no dressing of that particle-hole pair. In the alternative convention, using $\lambda$, the right-hand side is also amputated and this diagram would correspond simply to $\pm U$, as one is used to from the vertex $\gamma$. 

In terms of the Mandelstam variables~\cite{Mandelstam58} used in high energy physics, this is an s-channel diagram where the intermediate particle-hole pair transfers the frequency $\omega$, so that the resulting expression depends only on $\omega$ and not on $\nu$. In addition to depending on $\omega$ only, the term is also purely real since the susceptibilities $\chi^\rho$ (or the particle-hole bubble $B$) are real functions of (Matsubara) frequency in our situation, where the susceptibility is the temporal autocorrelation of the observable $\rho=n,S^z$ and satisfies $\chi(\tau)=\chi(-\tau)$.

Eq. \eqref{eq:lambda:unequalU} has contributions at order $U$ and $U^3$, which can be seen from the RPA expression for the susceptibilities, $\chi^{c/z} = \chi^{(0)}/\left(1 \mp U \chi^{(0)}\right)$.

The others channels contribute at order $U^2$,
\begin{align}
 L^{\dn\up}_{\nu\omega} |^\text{(b)} =& \frac{U^2}{\beta^2} \sum_{\nu'\nu_2} g_{\nu'}g_{\nu'+\omega} g_{\nu_2}g_{\nu_2+\nu'-\nu} \\
 L^{\dn\up}_{\nu\omega} |^\text{(b')} =& \frac{U^2}{\beta^2} \sum_{\nu'\nu_2} g_{\nu'}g_{\nu'+\omega} g_{\nu_2}g_{\nu+\nu'+\omega-\nu_2} 
\end{align}
These diagrams could also be formulated in terms of susceptibilities to incorporate higher-order corrections. However these would be susceptibilities in other channels, as can be seen in the diagrams in Table~\ref{table:diagrams}. Diagram (b) contains a propagating particle-hole pair of opposite spin, diagram (b') a particle-particle pair. These would be captured by the spin-flip susceptibility $\av{\left(c^\dagger_{\up} c^{\phantom{\dagger}}_{\dn}\right)_{\tau_2}\left( c^\dagger_{\dn} c^{\phantom{\dagger}}_{\up}\right)_{\tau_1}}$ and the particle-particle susceptibility $\av{\left(c^\dagger_{\up} c^\dagger_{\dn}\right)_{\tau_2} \left(c^{\phantom{\dagger}}_{\up} c^{\phantom{\dagger}}_{\dn}\right)_{\tau_1}}$ instead of the density-type susceptibilities $\av{\left(c^\dagger_{\up} c^{\phantom{\dagger}}_{\up}\right)_{\tau_2}\left( c^\dagger_{\dn} c^{\phantom{\dagger}}_{\dn}\right)_{\tau_1}}$ considered previously.

Based on the frequency argument in the respective bubbles, these diagrams are t-channel and u-channel respectively. These diagrams are not necessarily real, so that they provide the leading order for $\Im L^{\up\dn}$ since diagram (a) is purely real.

\subsubsection{$L^{\up\up}$}

Doing the same for the equal spin channel, we find two types of diagrams, in the s and the t channel, 
\begin{align}
 L^{\up\up}_{\nu\omega}|^\text{(c)} 
 &= -U^2 \frac{1}{\beta^2}\sum_{\nu',\nu_2}
g_{\nu'}g_{\nu'+\omega} g_{\nu_2} g_{\nu_2+\omega} \notag \\
 L^{\up\up}_{\nu\omega}|^\text{(d)} 
 &= +U^2 \frac{1}{\beta^2}\sum_{\nu',\nu_2}
g_{\nu'}g_{\nu'+\omega} g_{\nu_2}g_{\nu_2+\nu'-\nu} \notag\\ 
 L^{\up\up}_{\nu\omega} &\approx -1+L^{\up\up}_{\nu\omega}|^\text{(c)}+L^{\up\up}_{\nu\omega}|^\text{(d)}. \label{eq:L:upup}
\end{align}
Both can be reformulated in terms of the equal-spin susceptibility to include higher-order terms.
\begin{align}
  L^{\up\up}_{\nu\omega}|^\text{(c)} &= -U^2 B_{\omega} \chi^{\dn\dn}_\omega \notag \\
  L^{\up\up}_{\nu\omega}|^\text{(d)} &= +U^2 \frac{1}{\beta}\sum_{\nu'}
g_{\nu'}g_{\nu'+\omega} \chi^{\dn\dn}_{\nu'-\nu} \notag
\end{align}

The s-channel diagram is similar to the s-channel diagram for $L^{\up\dn}$ in several ways. First of all, it does not depend on $\nu$ in any way, since the only two lines that carry the frequency $\nu$ are amputated. Secondly, this contribution is real since both the susceptibility $\chi$ and the bubble $B$ are real.

The t-channel diagram, on the other hand, contains $\chi_{\nu-\nu'}$ which depends explicitly on $\nu$. This means that $L_{\nu\omega}^{\up\up}$ depends on $\nu$ at the second order in $U$. In addition, this contribution is not purely real in general. Based on this, we expect $\Im L$ and $\Im \lambda$ to scale as $U^2$ for small $U$. We will come back to real and imaginary parts of the vertex later.

\begin{figure}
 \includegraphics{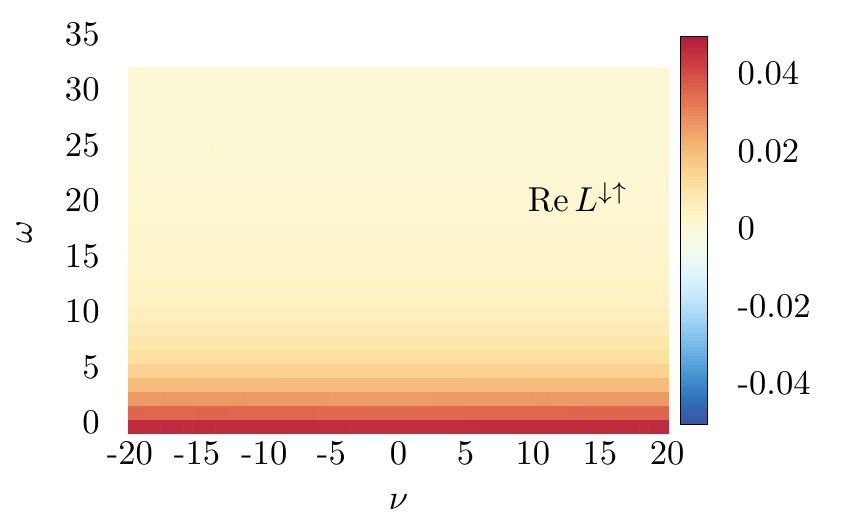}
 \includegraphics{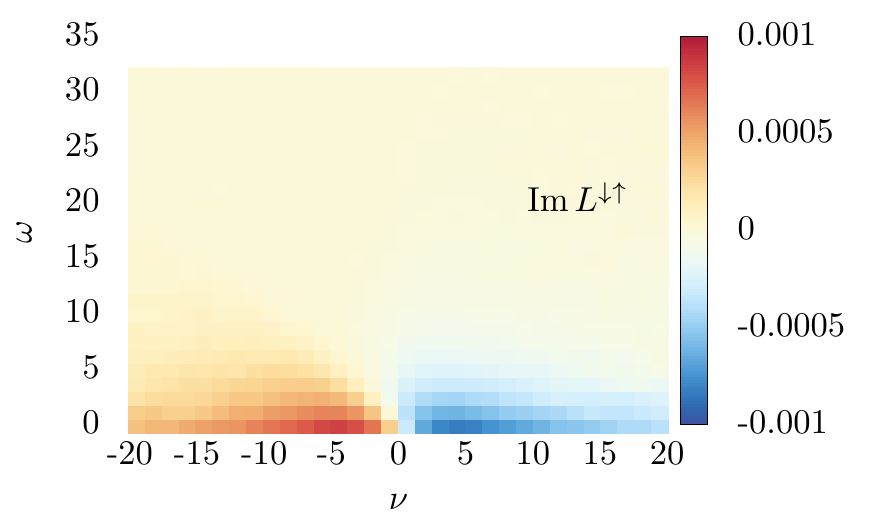}
 \includegraphics{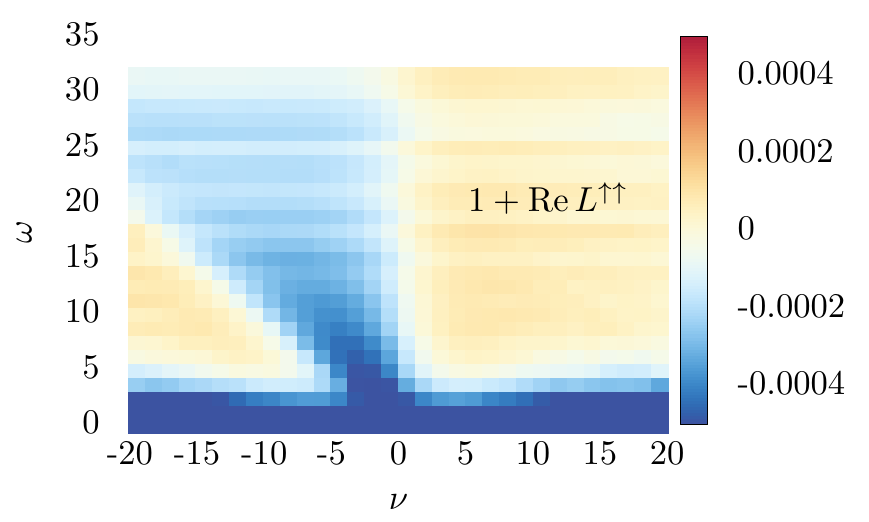}
 \includegraphics{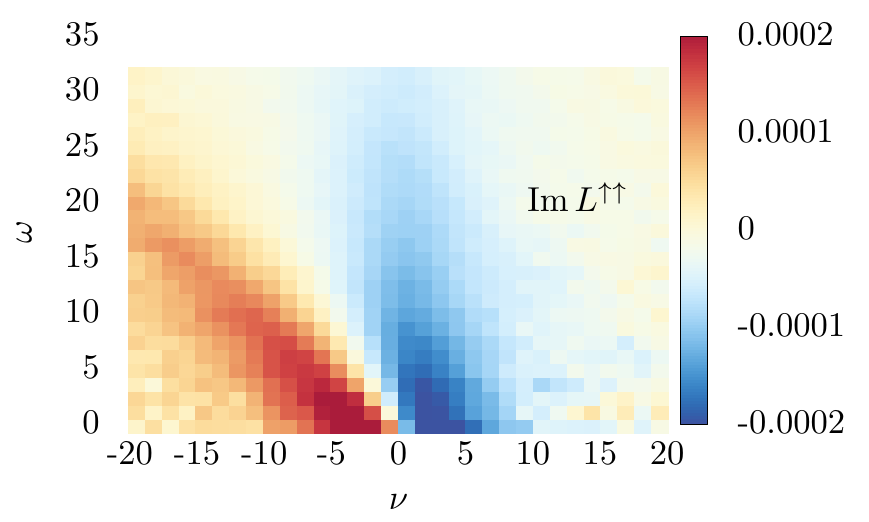}
 \caption{The fermion-boson vertex $L_{\nu\omega}$ at $U=0.4$, $\mu=-1$ and $\beta=5$, calculated within DMFT. Every pixel corresponds to a specific pair of Matsubara frequencies.}
 \label{fig:U0.4:colorplots}
\end{figure}

\begin{figure}
  \includegraphics{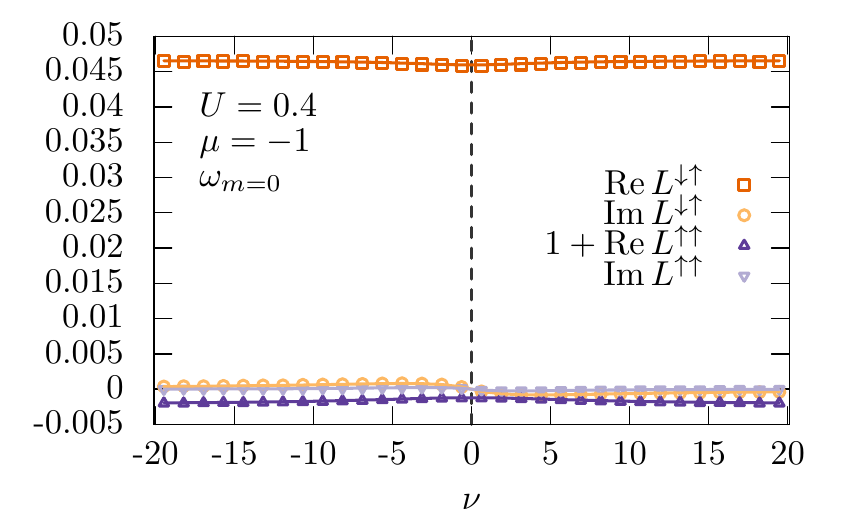}
  \includegraphics{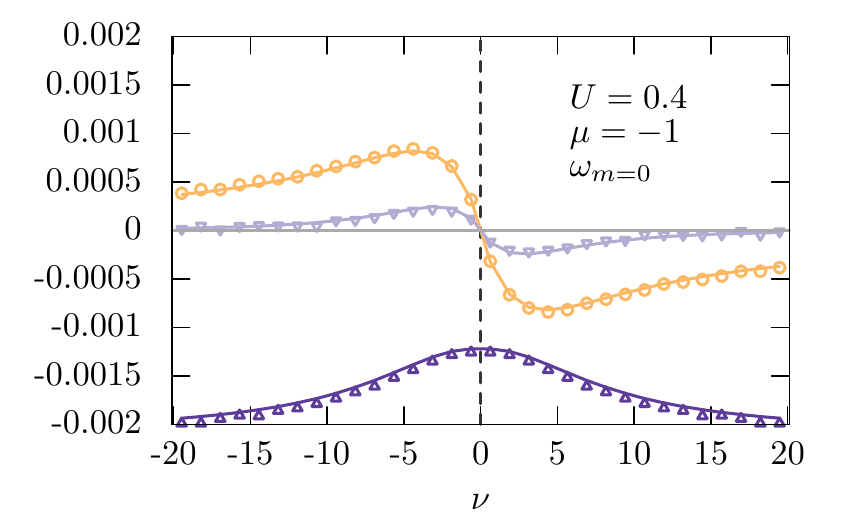}
  \includegraphics{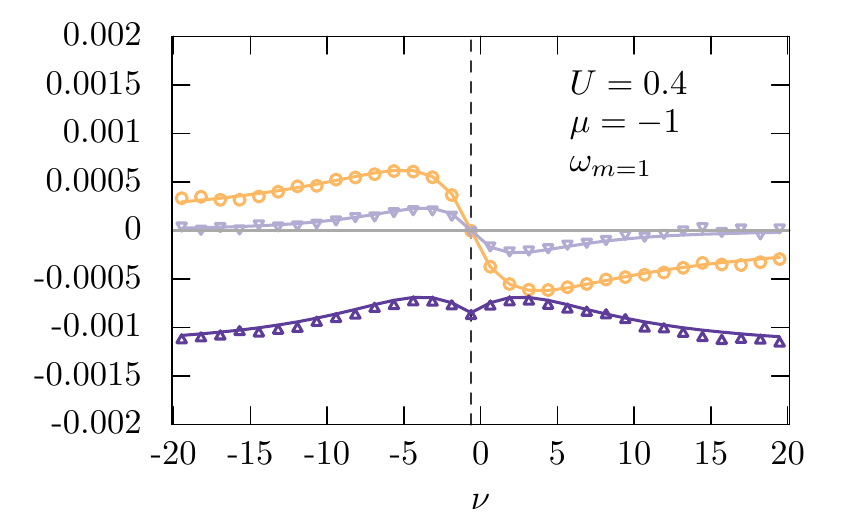}
  \includegraphics{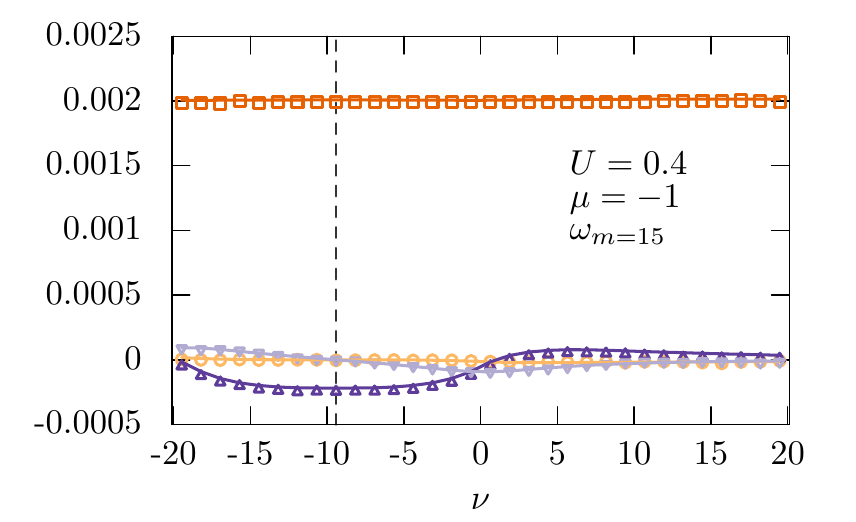}
  \caption{The fermion-boson vertex at $U=0.4$, $\mu=-1$ and $\beta=5$, cross-sections at several values of $\omega$ of the data in Fig.~\ref{fig:U0.4:colorplots}. The first two panels are both at $\omega=0$, the second one has a significantly smaller y axis for clarity. Solid lines are the perturbative, diagrammatic predictions. The vertical dashed gray line is the center point of the time-reversal symmetry relation, Eq.~\eqref{eq:timereversal}.}
  \label{fig:U0.4:plots}
 \end{figure}

Fig.~\ref{fig:U0.4:colorplots} shows numerical results at $U=0.4$, $\mu=-1$ and $\beta=5$, i.e., in the small $U$ regime. These results confirm the perturbative expectations: The largest contribution (after the trivial contribution -1 to $L^{\up\up}$) comes from diagram (a), is in the $L^{\dn\up}$ channel, purely real, independent of $\nu$ and decaying as a function of $\omega$. The perturbative expression do not just describe the qualitative features, they actually give a very good quantitative match at this small value of $U$, as can be seen by comparing the symbols and the solid lines in Fig.~\ref{fig:U0.4:plots}.
These two figures also show that the imaginary part and $\Re L^{\up\up}$ do depend on $\nu$, mostly in the range $\nu\in \left[-\omega,0\right]$.

\subsubsection{Particle-hole symmetry}

In this work, we study systems without particle-hole symmetry. However, our analytical results are derived for the general situation and are also applicable to particle-hole symmetric situations. Verifying that our diagrammatic expressions satisfy particle-hole symmetry is a useful consistency check. Particle-hole symmetry as it occurs for DMFT on bipartite lattices at half-filling means that~\cite{Tian97,Rohringer12} under the transformation $U\rightarrow -U$, all single-particle properties stay the same, at the two-particle level $S^z$ and charge fluctuations are interchanged (and $S^x$, $S^y$ are interchanged with the particle-particle fluctuations). Looking at Eq.~\eqref{eq:lambda:spinresolved}, this means that the following must hold:
\begin{align}
 L^{\up\up} \overset{U\rightarrow-U}{\Longleftrightarrow} \phantom{-}L^{\up\up} \\
 L^{\up\dn} \overset{U\rightarrow-U}{\Longleftrightarrow} -L^{\up\dn}  
\end{align}
The first relation is obviously satisfied by Eq.~\eqref{eq:L:upup}, since $L^{\up\up}$ depends only on $U^2$, $g$ and $\chi^{\dn\dn}$ and these are all invariant under particle-hole symmetry. For the second relation, we see that the first-order diagram for $L^{\up\dn}$ is indeed antisymmetric in $U$. It is a bit harder to see what happens with diagrams (b) and (b') at particle-hole symmetry. In fact, these two diagrams cancel in that case. Particle-hole symmetry implies $-g_\nu = g_{-\nu}$, i.e. the Green's function is purely imaginary, so that
\begin{align}
 L^{\dn\up}_{\nu\omega} |^\text{(b')} =& \frac{U^2}{\beta^2} \sum_{\nu'\nu_2} g_{\nu'}g_{\nu'+\omega} g_{\nu_2}g_{\nu+\nu'+\omega-\nu_2} \notag  \\
 =& (-1)^2 \frac{U^2}{\beta^2} \sum_{\nu'\nu_2} g_{-\nu'}g_{-\nu'-\omega} g_{\nu_2}g_{\nu+\nu'+\omega-\nu_2} \notag \\
 =& \frac{U^2}{\beta^2} \sum_{a\nu_2} g_{a+\omega}g_{a} g_{\nu_2}g_{\nu-a-\nu_2} \notag \\
 =& \frac{U^2}{\beta^2} \sum_{a\nu_2} g_{a}g_{a+\omega} g_{\nu_2}(-g_{a+\nu'-\nu}) \notag \\
 =& - L^{\dn\up}_{\nu\omega} |^\text{(b)},
\end{align}
where in the second line the summation variable was changed to $a=-\nu'-\omega$. 
This completes the proof that the expansion of $L$ satisfies particle-hole symmetry when the underlying $g$ and $\chi$ do.

\subsubsection{Time-reversal symmetry}

Another useful property of the fermion-boson vertex is the mirror symmetry  in the fermionic frequency axis~\cite{vanLoon14}, which comes from the time-reversal symmetry of the underlying action.
\begin{align}
(\lambda^{\rho,\sigma}_{\nu,\omega})^\ast
&= \lambda^{\rho,\sigma}_{-\nu-\omega,\omega}. \label{eq:timereversal}
\end{align}
In Ref.~\onlinecite{vanLoon14}, the fermion-boson vertex was only shown in a particle-hole symmetric system, where the vertex is purely real. The calculations performed here do not feature particle-hole symmetry so that both aspects, complex conjugation and frequency mirroring, of the symmetry are visible.

The time-reversal symmetry relation is properly satisfied by the diagrammatic expansion of $L$ shown in Table~\ref{table:diagrams}. The time-reversal symmetry relation interchanges diagrams (b) and (b'), the other diagrams are time-reversal symmetric by themselves.

In the figures, as for example in Fig.~\ref{fig:U0.4:plots}, the central point for the particle-hole symmetry, $\nu=-\omega/2$, is indicated by a vertical gray line for $\omega \neq 0$. The (anti-)symmetry of the vertex with respect to this frequency is clearly visible.

\section{Ward identity}

In the previous section, we used a perturbative expansion to predict the $L$ based on simpler objects. Such a perturbative approach is only suited to weakly interacting systems, whereas DMFT and its expansions are specifically aimed at correlated systems. In this section, we use the Ward identity~\cite{Ward50, Takahashi1957,Hertz73} to derive relations between $L$ and simpler objects. Since the Ward identity is an exact relation in quantum field theory that holds regardless of the interaction strength, this approach is useful even for strongly correlated systems. The Ward identity is a manifestation of the continuity equation for the density operator, $\partial_\tau\rho+[\rho,H_\text{AIM}]=0$. A recent discussion of the role of the Ward identity in DMFT can be found in Ref.~\onlinecite{Krien17}.

\begin{figure}
  \includegraphics{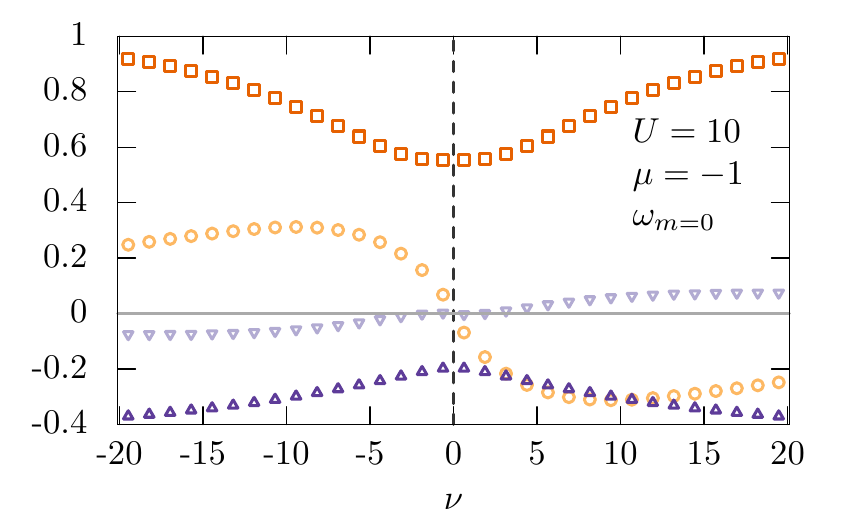}
  \includegraphics{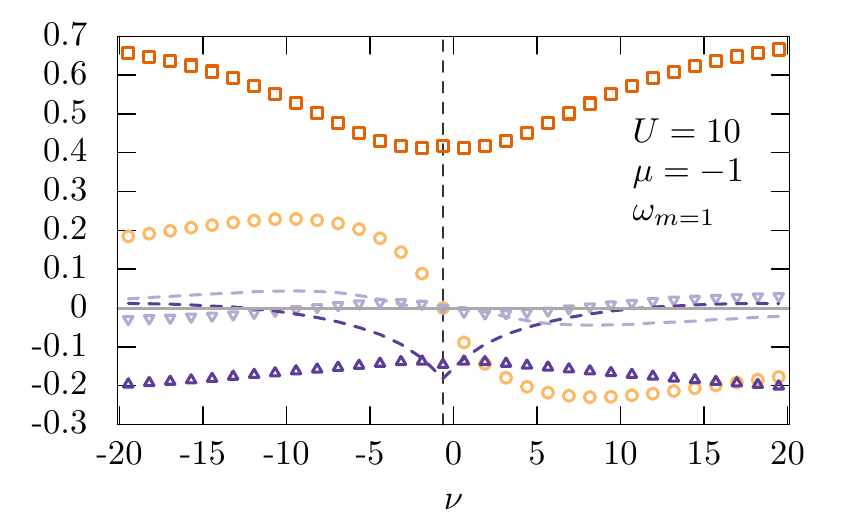}
  \includegraphics{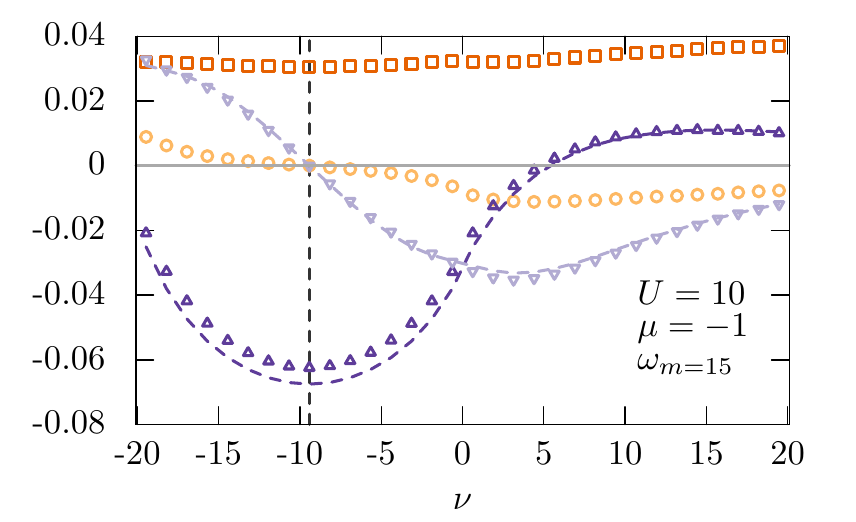}
  \caption{The fermion-boson vertex at $U=10$, $\mu=-1$ and $\beta=5$, cross-sections at two values of $\omega$. Colors are the same as in Fig.~\ref{fig:U0.4:plots}. Dashed lines are the lowest order contribution to the Ward identity, Eq.~\eqref{eq:ward:approx} at $\omega \neq 0$. }
  \label{fig:U10:plots}
 \end{figure}

For the fermion-boson vertex at a finite frequency $\omega$, the Ward identity may be written as (see Appendix~\ref{app:ward})
\begin{align}
    L^{\up\up}_{\nu\omega}=-&1+\frac{\Sigma_{\nu+\omega}-\Sigma_\nu}{\imath\omega} \notag \\
    +&\frac{1}{\beta}\sum_{\nu'}\frac{\Delta_{\nu'+\omega}-\Delta_{\nu'}}{\imath\omega}\gamma^{\up\up}_{\nu\nu'\omega}g_{\nu'}g_{\nu'+\omega},\label{eq:wardid} \\
    L^{\up\dn}_{\nu\omega}=    +&\frac{1}{\beta}\sum_{\nu'}\frac{\Delta_{\nu'+\omega}-\Delta_{\nu'}}{\imath\omega}\gamma^{\up\dn}_{\nu\nu'\omega}g_{\nu'}g_{\nu'+\omega},\notag 
\end{align}
where we remind the reader that $\Delta_{\nu}$ is the hybridization function that enters the impurity action, Eq.~\eqref{eq:Simp}. We note that the right-hand side of the Ward identity also contains a two-particle correlation function, $\gamma$. In that sense, no simplification has been achieved so far. However, this identity immediately suggests as useful approximation for $L^{\up\up}$,
\begin{align}
 L^{\up\up} \approx -1+\frac{\Sigma_{\nu+\omega}-\Sigma_\nu}{\imath\omega}. \label{eq:ward:approx}
\end{align}
Fig.~\ref{fig:U10:plots} shows the fermion-boson vertex at a sizeable value of $U$, $U=10$, for $\mu=-1$ which corresponds to $\av{n}\approx 0.43$. The dashed lines show this approximation (for $\omega\neq 0$), the symbols show the numerical results. We observe that this simple approximation works well at large $\omega$, where it captures both the magnitude and the frequency structure rather accurately. On the other hand, for smaller $\omega$ there are significant deviations coming from the vertex corrections in Eq.~\eqref{eq:wardid}. In particular, ignoring the vertex corrections gives $L^{\up\dn}=0$, or in other words, $L^{n} = L^{z}$. A more sophisticated and useful approximation is discussed in Appendix~\ref{app:asymptote}.

It is also immediately clear that Eq.~\eqref{eq:ward:approx} works well in the limit $U\rightarrow 0$ since $\gamma^{\up\dn}\propto U$ and $\gamma^{\up\up}\propto U^2$. Note also that although there are linear in $U$ contributions to the self-energy (the Hartree diagram, the Fock diagram vanishes since it couples fermions with the same spin), these are independent of frequency so the second term in Eq.~\eqref{eq:wardid} also contributes at order $U^2$, as expected from the diagrammatic expressions given earlier. In fact, as shown in Appendix~\ref{app:deltaU0}, diagram (a) is recovered from the Ward identity in the weakly interacting limit where $\Delta$ and $g$ take their non-interacting values.

With respect to the discussion in Sec.~\ref{sec:physicalcontent}, it is important to note that the Ward identity presented here is based on a continuity equation, not on a conservation relation. The total charge and magnetization of the impurity are not conserved quantities, since fermions can move from and onto the impurity from the bath, which acts as a ``source'' and ``sink''. These movements are responsible for the appearance of the term containing $\Delta$ and $\gamma$. In fact, in the atomic limit which is discussed below, Eq.~\eqref{eq:ward:approx} is exact. This means that the approximation correctly describes both limits of the Hubbard model.

\section{Sum rule}

As was discussed above, the fermion-boson vertex is obtained from the fermion-fermion vertex by ``tying together two fermion lines into a boson''. This can be done once more, resulting in a sum rule relating $\lambda$ to the susceptibility $\chi$~\cite{Stepanov16}. In terms of $L$,
\begin{align}
 \frac{1}{\beta}\sum_{\nu} g_{\nu}g_{\nu+\omega} L^{\up\up}_{\nu\omega} =& \av{n_{\up,\omega} n_{\up,\omega}} - \av{n_\up}\av{n_\up}\delta_\omega \notag \\ =& -\chi_{\omega}^{\up\up}  \text{\quad\quad (Sum $L^{\up\up}$)}
\notag \\
 \frac{1}{\beta}\sum_{\nu} g_{\nu}g_{\nu+\omega} L^{\up\dn}_{\nu\omega} =& \av{n_{\up,\omega} n_{\dn,\omega}} - \av{n_\up}\av{n_\dn}\delta_\omega \notag \\ =& -\chi_{\omega}^{\up\dn}
 \text{\quad\quad (Sum $L^{\up\dn}$)}
 \label{eq:sumrule1}
\end{align}

These relations follow straightforwardly from Eq.~\eqref{eq:lambda:spinresolved}. The susceptibility on the right-hand side of these equations is purely real, the left-hand side is also real due to time-reversal symmetry, Eq.~\eqref{eq:timereversal}.

This sum rule is indeed satisfied numerically in our computations. However, the convergence with respect to the summation variable $\nu$ is relatively slow, as illustrated below.

The Ward identity can also be written as a sum rule, as shown in App.~\ref{app:ward},
\begin{align}
 \frac{1}{\beta} \sum_{\nu} \frac{\Delta_{\nu+\omega}-\Delta_\nu}{\imath\omega}  g_{\nu} g_{\nu+\omega} L^{\up\up}_{\nu\omega} =& -\chi^{\up\up}_\omega 
 \text{\quad (Ward $L^{\up\up}$)}
 \notag \\
 \frac{1}{\beta} \sum_{\nu} \frac{\Delta_{\nu+\omega}-\Delta_\nu}{\imath\omega} g_{\nu} g_{\nu+\omega} L^{\up\dn}_{\nu\omega} =& -\chi^{\up\dn}_\omega 
 \text{\quad (Ward $L^{\up\dn}$)}
 \label{eq:sumruleward}
\end{align}
Comparing this sum rule to Eq.~\ref{eq:sumrule1}, we see that inserting the non-trivial factor $(\Delta_{\nu+\omega}-\Delta_{\nu})/\imath\omega$ does not change the outcome of the sum. This is similar to the result of Appendix~\ref{app:deltaU0}.
The new sum rule is useful, however, since the convergence of Eq.~\eqref{eq:sumruleward} is much better than that of Eq.~\eqref{eq:sumrule1}, since the additional factor typically decays as a function of $\nu$. 

\begin{table}
 \begin{tabular}{l|cccc}
 $m$
   & \eqref{eq:sumrule1} $L^{\up\up}$
   & \eqref{eq:sumrule1} $L^{\up\dn}$
   & \eqref{eq:sumruleward} $L^{\up\up}$
   & \eqref{eq:sumruleward} $L^{\up\dn}$
   \\
 \hline
  0  & 0.794  & 0.673 & N/A & N/A \\
  1  & 0.750  & 0.580 & 0.999 & 0.999 \\
  15 & -4.36 & -5.05 & 0.899  & 0.906 \\
 \end{tabular}

 \caption{Partial fulfillment of the sum rules \eqref{eq:sumrule1} and \eqref{eq:sumruleward} when the summation over fermionic frequency $\nu$ is restricted to $\nu_n \in [-20,20 ]$ (32 Matsubara frequencies). This is the same frequency range as shown in Fig.~\ref{fig:U10:plots}, these numbers also correspond to the same parameters, $U=10$, $\mu=-1$ and $\beta=5$. Complete fulfillment of the sum rule corresponds to +1.0 in this table. The first column indicates the bosonic Matsubara frequency $\omega_m$. See also Fig.~\ref{fig:U10:sumrule}, which shows the contribution to the sum rule per fermionic frequency. }
 \label{tab:sumrules:U10}
\end{table}

\begin{figure}
  \includegraphics{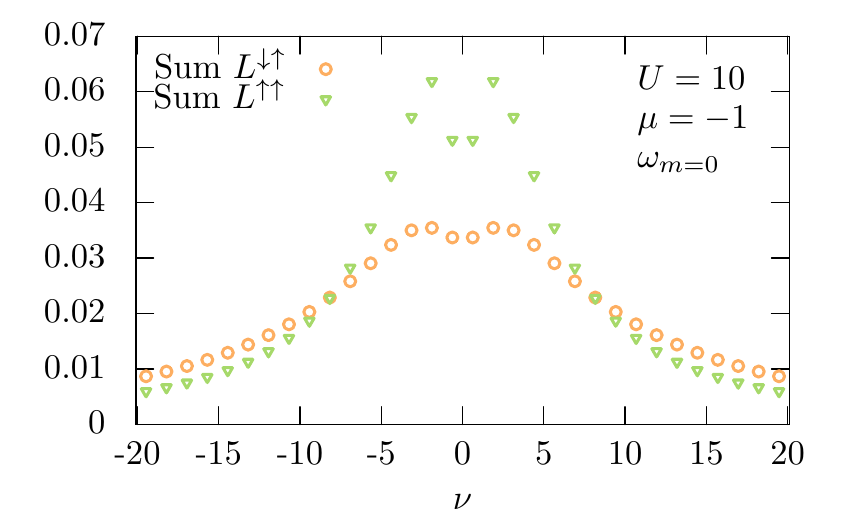}
  \includegraphics{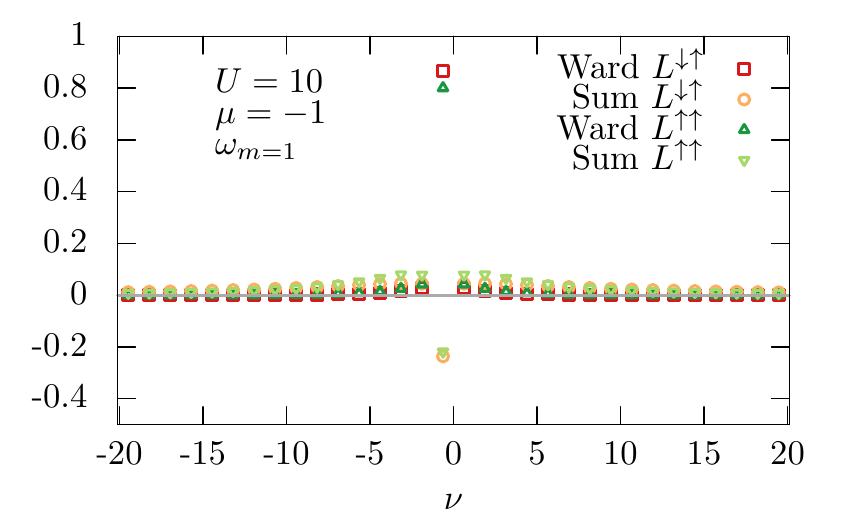}
  \includegraphics{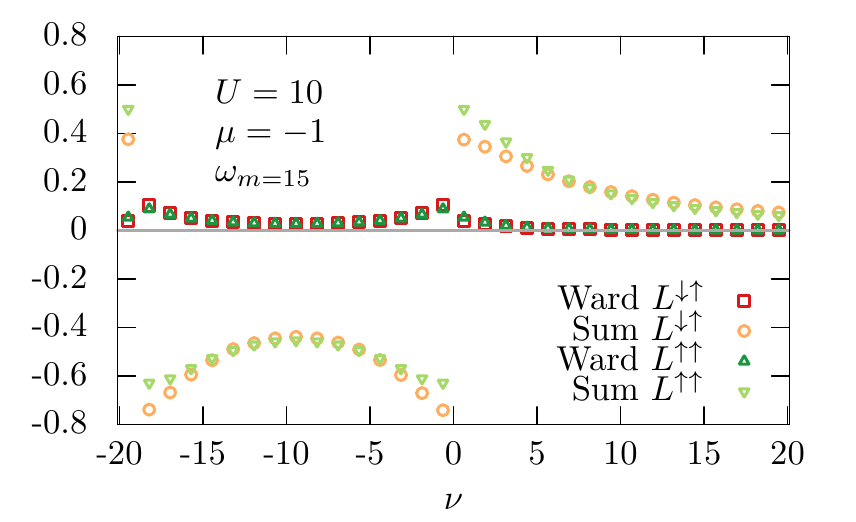}
  \caption{
  Fractional contribution to the sum rules \eqref{eq:sumrule1} and \eqref{eq:sumruleward} as a function of $\nu$, i.e. for Sum $L^{\up\up}$ this graph shows $-\frac{1}{\beta} \frac{1}{\chi_\omega} g_{\nu}g_{\nu+\omega} L^{\up\up}_{\nu\omega}$. The parameters are $U=10$, $\mu=-1$ and $\beta=5$, for three values of $\omega$, see also Fig.~\ref{fig:U10:plots} where the fermion-boson vertex at the same parameters is shown. Table~\ref{tab:sumrules:U10} contains these quantities summed over fermionic frequency. }
  \label{fig:U10:sumrule}
\end{figure}

We illustrate the sum rules in Table \ref{tab:sumrules:U10} and Fig.~\ref{fig:U10:sumrule}, corresponding to the vertex of Fig.~\ref{fig:U10:plots}. For clarity, the results have been divided by minus the susceptibility, so that the sum should equal $+1$.  Only the real part of the sum rules is shown, the imaginary part is fulfilled as a consequence of time-reversal symmetry. The Ward identity sum rule is only applicable at $\omega\neq 0$.

Several observations can be made from these results. First of all, the convergence of the Ward sum rule is much better, with most of the sum rule restricted to the range $(-\omega,0)$. For the normal sum rule, this range actually contributes to the sum rule with the wrong sign, so that the remaining frequencies need to compensate to eventually fulfill the sum rule. At $\omega_{m=15}$, this compensation is so incomplete that the partial sum over the frequency range shown here is negative. The range $\nu\in (-\omega,0)$ describes exactly the fermions with $\nu<0$ and $\nu+\omega>0$, i.e. fermions that are excited from below to above the real axis by the external field.

Secondly, we see that the contributions to the sum rule are more spread out over $\nu$ for larger $\omega$. This results in a decreasing fractional fulfillment of the sum rules for large $\omega$, a larger $\nu$ range is needed to fully capture the sum rules.

Finally, the fulfillment of the sum rules for $L^{\up\up}$ is better than those for $L^{\up\dn}$. This is consistent with the results of Fig.~\ref{fig:U10:plots}, where $\Re L^{\up\dn}$ (dark orange) actually increases as a function of $\nu$, whereas $\Re L^{\up\up}$ is dominated by the constant trivial contribution $-1$. To rephrase, the difference in the charge and magnetic response occurs at relatively large fermionic Matsubara frequency (for fixed bosonic frequency, i.e., frequency of the applied field). On the other hand, the difference between charge and magnetic response vanishes for large bosonic Matsubara frequency~\cite{Krien17}. 

These three observations hold not only for the specific parameters shown here, we have verified them over a large part of phase space. 

The sum rules essentially tell how much of the total response is contained in a fermionic frequency window. With an eye on diagrammatic extensions of DMFT, where the fermion-boson vertex is usually only calculated for a finite number of fermionic frequencies, the sum rule can provide a guide to what frequency cutoff should be used. However, in this respect it is important to remember that the dual boson method~\cite{Rubtsov12,vanLoon14} combines the fermion-boson vertex with a so-called dual Green's function that decays as $(i\nu)^{-2}$ instead of the $(i\nu)^{-1}$ decay of $g$, so convergence of dual boson calculations is likely to be better than the sum rule suggests.

\section{Atomic limit}

Additional insight into the physics of the Hubbard model can be gained by studying the atomic limit~\cite{Hafermann09,Kozik15,Ayral15,Ayral16,Wentzell16,Wu17,Gunnarsson17,Tarantino17,Thunstrom18}. This limit can be understood either as the $t\rightarrow 0$ limit of a Hubbard lattice model, or simply as a single-site Hubbard model. Since the Hilbert space of a single atom is small, the exact eigenstates can be found and used to calculate all observables.

In the DMFT spirit, the atomic limit corresponds to setting $\Delta=0$. Looking at the Ward identity, \eqref{eq:wardid}, the terms containing $\gamma$ vanish and for $\omega>0$,
\begin{align}
L_{\text{atomic}}^{\up\dn}=& 0 \notag \\
L_{\text{atomic}}^{\up\up} =& -1+\frac{\Sigma_{\nu+\omega}-\Sigma_\nu}{\imath\omega} \label{eq:ward:atomic} \\ =&-1+\frac{U^2}{4\nu(\nu+\omega)}. \notag
\end{align}
Here, the second equation holds at half-filling and uses the well known expression for the self-energy in the half-filled Hubbard atom $\Sigma_\nu = U^2/4i\nu$. This expression clearly satisfies the time-reversal symmetry and particle-hole symmetry requirements. This results has previously been derived~\cite{Ayral15,Ayral16} based on the Lehmann representation instead of the Ward identity. Alternatively, one can also verify that summing over the second fermionic frequency $\nu'$ in the expression for the generalized susceptibility of the atomic limit~\cite{Thunstrom18} gives the same result. The Ward identity approach is more elegant and efficient.

The qualitative difference between the atomic limit and the impurity embedded in a bath is that the fermion number $n_\sigma$ is a conserved quantity in the atomic limit, the fermion cannot go anywhere. In that sense, Eqs.~\eqref{eq:ward:atomic} are identical to the $q=0$ Ward identity for lattice models, which also corresponds to conserved quantities.
That $L^{\up\dn}$ is zero can be understood when interpreting $L$ as a response function. A finite frequency field which only works on the $\up$ fermions will change their spectrum but not the total particle number, so the $\dn$ electrons will not feel anything.

It is also useful to have a look at the sum rule obtained from the Ward identity, Eq.~\eqref{eq:sumruleward}. The left-hand side of this equation is zero for $\omega\neq 0$. The right-hand side contains the susceptibility of the Hubbard atom. This susceptibility is also zero at finite frequency, since the total charge of either spin in the atom is conserved, the fermions cannot move. The atomic limit illustrates once more why using $L$ instead of $\lambda$ can be useful, the definition of $\lambda$ includes a division by $\chi$ which is zero, so that $\lambda$ is divergent in the atomic limit.

\section{Real energies}

So far, we have studied the fermion-boson vertex using Matsubara frequencies. An analysis using real frequencies is also possible and can be useful for a physical interpretation. A Lehmann representation for three-point correlation functions exists~\cite{Oguri01,Tagliavini18}, expressing the correlation function in the complex plane in terms of the exact eigenstates of the model. For practical purposes, one can consider analytical continuation of the vertex from the Matsubara axis to real frequencies. This continuation is a bit more involved than that of the single-particle Green's function, since the fermion-boson vertex has two frequency arguments. We use the notation $\nu\rightarrow E+i\eta$ and $\omega\rightarrow \Omega+i\zeta$.

For the special case $\omega_{m=0}=0$, no analytical continuation in $\omega$ is needed, since this frequency is both a real energy and a Matsubara frequency. 
In that case, we can simply use the regular toolbox for analytical continuation of single-particle Green's functions to do the continuation $\nu\rightarrow E+i\eta$.

Note that here, we have taken the limit $\omega_{m=0} \rightarrow 0$ first, before doing the analytical continuation. In particular, this means that $\abs{E}>\Omega=0$. This is (potentially) different from the limit $\lim_{\Omega\rightarrow 0} \lim_{E\rightarrow 0}$, i.e., with $\abs{E} < \abs{\Omega}$. 
This is similar to the static and dynamic limits of Fermi liquid theory, when energy and momentum go to zero, as discussed, for example, in \cite{Krien18}.

A further question concerns the normalization or sum rules. The imaginary part of the single-particle Green's function $g(E+i\eta)$ corresponds to the spectral function which is normalized to unity, and combining the spectral function with the Fermi function gives the particle density,
\begin{alignat*}{2}
 -\frac{1}{\pi} \int dE \,  \Im g(E)& &&= 1 \\
 -\frac{1}{\pi} \int dE \,  \Im g(E)& f(E) &&= \av{n}. 
\end{alignat*}
Here $f(E)$ is the Fermi distribution function.
Accordingly, we can immediately derive sum rules for $ggL$ by interpreting it as a response function, Eq.~\eqref{eq:L:abstract2}. First, suppressing the infinitisimal imaginary shift $+i\eta$ for notational convenience, and restricting ourselves to $\Omega=0$, we find 
\begin{align}
  \int dE \, \Im g_E g_{E} L_{E,\Omega=0}=& \int dE \, \frac{d}{dh} \Im g(E) \notag \\
  =& \frac{d}{dh} \left(-\pi\right) \notag \\
  =& 0, \label{eq:realenergy1}
\end{align}
which tells us that external fields do not change the total spectral weight. Secondly, 
\begin{align}
  \int dE \, f(E) \Im g^\up_E g^\up_{E} L^z_{E,\Omega=0}=& \int dE \, f(E) \frac{d}{dh} \Im g^\up(E) \notag \\
  =& -\pi \frac{d}{dh} \av{n_\up} \notag \\
  =& \frac{\pi}{2} \chi_{\Omega=0}, \label{eq:realenergy2}
\end{align}
which is simply the real energy equivalent~\footnote{The fermion-boson vertex of the impurity model corresponds to the response to an external field \emph{while keeping the impurity model constant}, as also discussed in footnote~\onlinecite{endnote108}.} of the sum rule Eq.~\eqref{eq:sumrule1}. Note that the factor of two comes from the fact that the magnetic susceptibility describes the change in magnetization, and in a paramagnetic system $d(\av{n_\up}-\av{n_\dn})/dh=2d\av{n_\up}/dh$. A similar sum rule can be derived for the density channel. The second sum rule can also be formulated in terms of $\lambda$ instead of $L$, in which case the right-hand side will simply be $\pi/2$. Note that in this case, since both sum rules are independent of $\Omega$, they actually also hold for Matsubara frequencies $\omega$, i.e. when only continuation in the fermionic frequency is performed.

Together, since $f(E)>0$, these two relations essentially imply that $gg \lambda$ contains contributions with opposite sign as a function of $E$, so that the total integral is zero whereas the weighted integral is finite. These opposite sign contributions happen at $E>0$, i.e. above the Fermi level, so that they do not fully enter Eq.~\eqref{eq:realenergy2}.

\begin{figure}
 \includegraphics[width=\columnwidth]{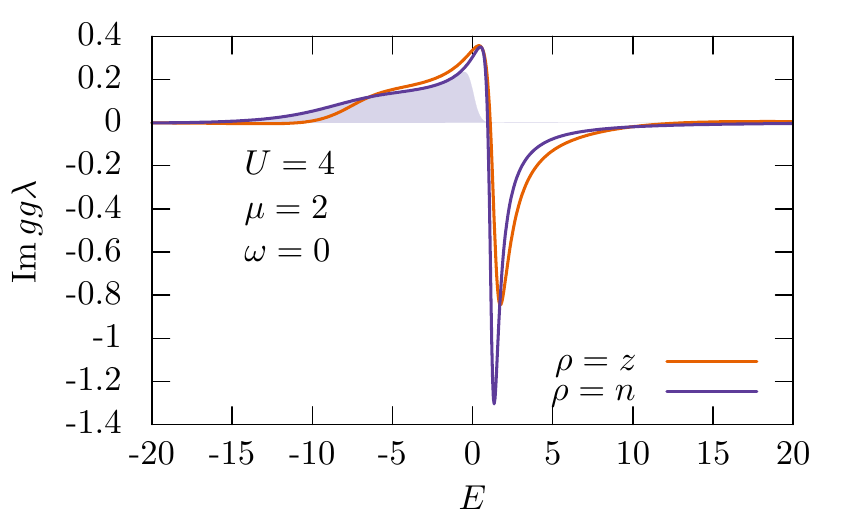}
 \caption{Vertex at $U=4$ and $\mu=2$ ($\av{n}\approx 0.9$) for $\omega=0$ after analytical continuation to real energies, using Pad\'e with $N_w=15$ and $\eta=0.3$. The lines indicate $gg\lambda$ which should integrate to zero according to Eq.~\eqref{eq:realenergy1}, the filled curve shows ${gg\lambda^n\cdot f(E)}$ which should integrate to $\pi/2$ according to Eq.~\eqref{eq:realenergy2}. In this case, the numerical deviation from these sum rules is less than 3\%.}
 \label{fig:U4:realenergy}
\end{figure}

The fact that $ggL$ is not positive definite hinders some forms of analytical continuation, although Eq.~\eqref{eq:realenergy2} could still be used as a normalizing factor. Here, to give only an impression of the real energy structure, we use Pad\'e approximants~\cite{Vidberg77} to perform the analytical continuation in the fermionic Matsubara frequency. In Fig.~\ref{fig:U4:realenergy} we show an example of such an analytical continuation.

In general, analytical continuation of data obtained from Monte Carlo simulations is a hard problem (see, e.g., Refs.~\cite{julich2013mishchenko,Schott16} for recent overviews). For this reason, here we have restricted ourselves to a moderate interaction strength where we were able to obtain the Matsubara axis data with sufficient accuracy to perform analytical continuation. Figure~\ref{fig:U4:realenergy} corresponds to a 'best case' for the continuation, we found significant continuation artifacts at many other parameters. A more systematic investigation of the real energy structure as a function of physical parameters would require both higher quality data and more sophisticated continuation procedures, possibly drawing inspiration from recent works on analytical continuation~\cite{Otsuki17,Kraberger17,Yoon18}.

Figure~\ref{fig:U4:realenergy} nicely illustrates how the simultaneous fulfillment of Eqs.~\eqref{eq:realenergy1} and \eqref{eq:realenergy2} occurs: Negative contributions at $E>0$ (above the Fermi level) ensure that the total spectral weight remains constant, the external field simply moves spectral weight from above to below the Fermi level. These states were initially (almost) empty and are now (mostly) filled, so that the total density does change. 

\section{Metal-Insulator transition}

One of the main motivations for DMFT comes from its ability to describe the Mott-Hubbard transtion. Initially, this transition was mainly studied on the single-particle level~\cite{Jarrell92,Rozenberg92,Rozenberg94,Georges96}, in terms of the vanishing quasi-particle weight $Z$ and the density of states. Later, it turned out that there are also very characteristic features of the Mott-Hubbard transition on the two-particle level~\cite{Rohringer12,Schafer13,vanLoon14b}. In this section, we discuss what happens to the fermion-boson vertex near the Mott transition and relate this to what happens on the single-particle level using the Ward identity.
We focus on the doping-driven transition, i.e., we keep temperature and interaction strength constant and change the chemical potential, the transition will be visible around half-filling (for sufficiently large $U$). Strictly speaking, the system is insulating only exactly at half-filling. The transition is studied by approaching this insulating state from above and from below (in terms of the density), which shows divergences and differences between taking the limit from above and from below, two signs of a phase transition.

\begin{figure}
 \includegraphics[width=\columnwidth]{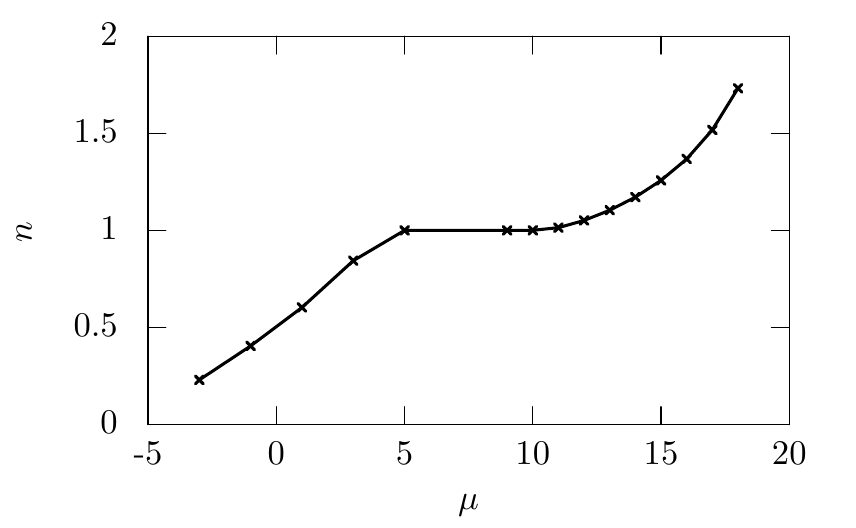}
 \caption{Density versus chemical potential at $U=16$ and $\beta=5$.}
 \label{fig:U16:nvsmu}
\end{figure}

The essential role of the fermion-boson vertex close to the metal-insulator transition is clear from Eq.~\eqref{eq:L:abstract}. The insulator has a charge gap, which means that the charge response $dn/d\mu$ is suppressed~\footnote{Note that the situation is different at a second-order Mott transition where the compressibility is actually divergent~\cite{Furukawa91,Kotliar02} instead of vanishing.}, the $n(\mu)$ curve of Fig.~\ref{fig:U16:nvsmu} goes almost completely flat. At the same time, the spin response is not suppressed. In the expression for the response, only $L$ depends on the channel, so it is clear that there must be large difference in the two channels of $L$. In particular, $L^n$ must suppress the finite value of $gg$ to a very small value of $dn/d\mu$~\footnote{There is a subtle point in this argument: The response functions of the DMFT solution for the lattice and of the auxiliary impurity model are not the same~\cite{vanLoon15,vanLoon16}. Essentially, the $dn/d\mu$ of the impurity model denotes the derivative with fixed hybridization $\Delta$ whereas the $dn/d\mu$ in DMFT also includes the effect of the hybridization changing as a function of $\mu$. The latter corresponds to the derivative of Fig.~\ref{fig:U16:nvsmu}. This distinction leads to quantitative differences, however the overall physics of a strongly suppressed compressibility is the same. The vertices shown in this manuscript all correspond to the impurity model.}. 

We start by looking at $U=16$, which is sufficiently large to be deep inside the insulating phase at half-filling.
In Fig.~\ref{fig:U16:nvsmu}, we show how the density evolves as a function of $\mu$. Half-filling occurs between $\mu=5$ and $\mu=9$. In this region, the compressibility $dn/d\mu$ (the slope of the curve) is very small, as expected close to an insulating phase. The numerical stability of the DMFT self-consistency cycle, which is subject to Monte Carlo errors, deteriorates when getting very close to half-filling. Our CT-HYB solver~\cite{Hafermann13} requires $\Delta_\tau<0$ for all $\tau$ and $-\Delta_{\tau=\beta/2}$ becomes very small close to the insulating phase, so that Monte Carlo noise can break this condition. The results shown here are sufficiently close, $\abs{\delta n} < 10^{-3}$ for $\mu=5$ and $\mu=9$, to half-filling to make meaningful statements about the transition.

\begin{figure}
  \includegraphics{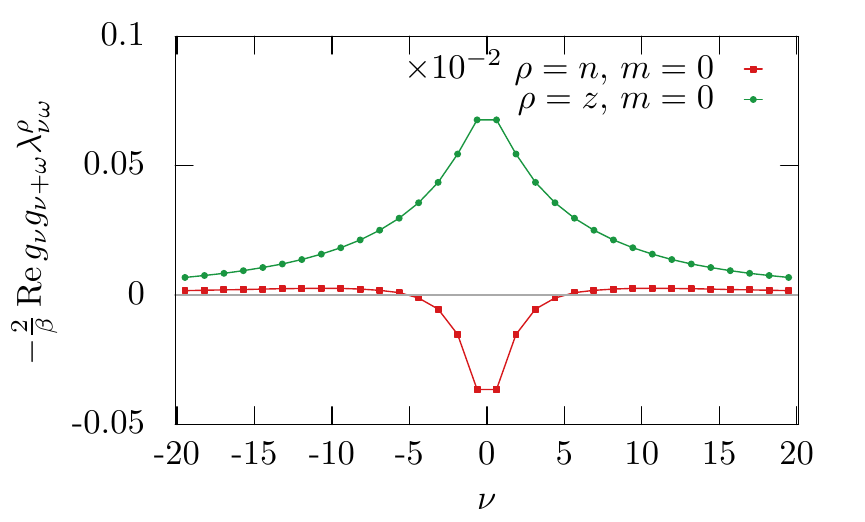}
  \includegraphics{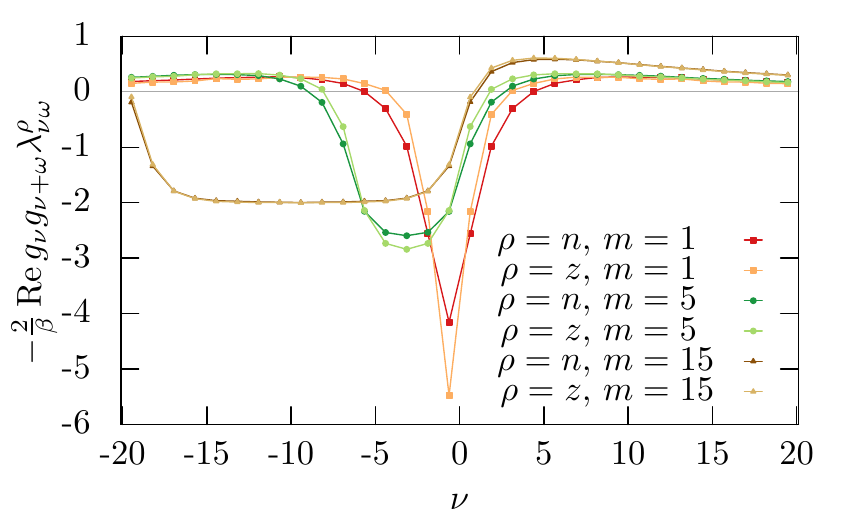}
  \caption{
  The combination $-2/\beta \cdot \Re g_\nu g_{\nu+\omega} \lambda^\rho_{\nu\omega} $, shown for $U=16$, $\mu=5$ and $\beta=5$, for several values of $\omega_m$. The sum over frequencies is equal to unity. The red line (charge channel) in the upper panel is scaled by $10^{-2}$ to be comparable in magnitude to the magnetic channel. }
  \label{fig:U16:sumrule}
\end{figure}

Close to half-filling, the susceptibility $\chi$ at $\omega=0$ is very different in the two channels, $(dn/d\mu)^\text{imp}=\chi_{\omega=0}^n\approx 0$ whereas $(dm/dh)^\text{imp}=\chi^z_{\omega=0}$ stays finite. 
As explained in Sec.~\ref{sec:physicalcontent}, the fermion-boson vertex describes how the individual fermionic levels contribute to these susceptibilities. The combination $-2/\beta \cdot g_\nu g_{\nu+\omega} \lambda_{\nu\omega}$ is normalized to unity, together the fermionic levels give the entire response. 

Figure~\ref{fig:U16:sumrule} shows this quantity just below half-filling. In the charge channel, we see that the fermions at small $\nu$ give large negative contributions that are eventually compensated by positive contributions at large $\nu$. 
This important role of negative contributions to $dn/d\mu$ is reminiscent of the eigenvalue analysis of Ref.~\onlinecite{Gunnarsson17}.
The charge response had to be rescaled to be comparable to the magnetic response. The latter has not changed much in shape compared to lower interaction strengths, the largest contribution to $dm/dh$ comes from small $\nu$, decaying monotonously for large $\nu$.
 
For $\omega \neq 0$, we find negative small $\nu$ contributions of significant magnitude in both channels. For large $\omega$, the vertex is independent of the channel. This is consistent with the fact that at large $\omega$, both susceptibilities share the same asymptote~\cite{Krien17}. The insulating state is characterized by the difference in charge and spin response at low (zero) frequency, however charge and spin excitations are indistinguishable at high frequencies. 

\begin{figure}
 \includegraphics{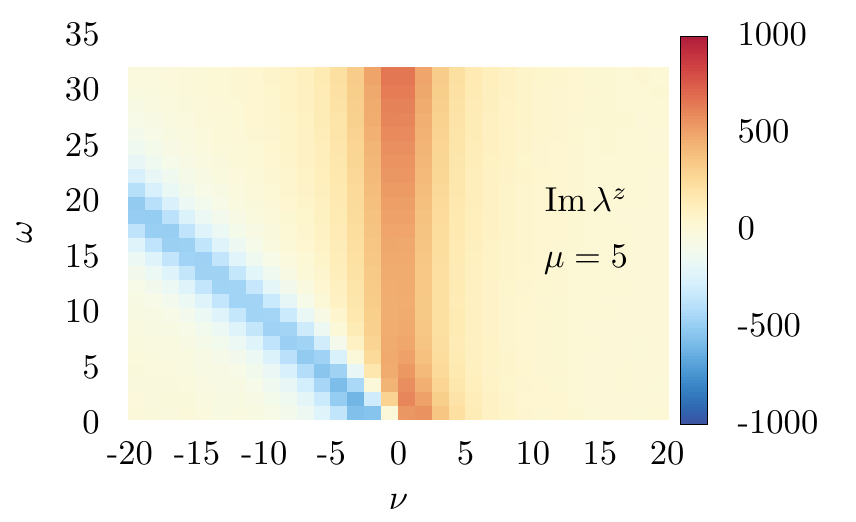}
 \includegraphics{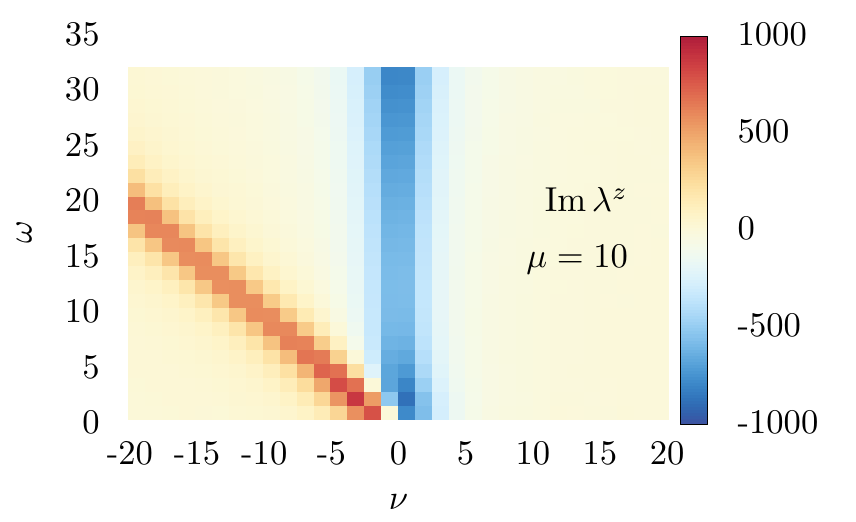}
 \caption{Fermion-boson vertex, at $U=16$, for $\mu=5$ (below half-filling) and $\mu=10$ (above half-filling). Only $\omega>0$ is shown.}
 \label{fig:U16:colorplots:im} 
\end{figure}

Figure~\ref{fig:U16:colorplots:im} shows the imaginary part of $\lambda$ for $\omega>0$, both below and above half-filling. The imaginary part is large in magnitude and changes sign at half-filling, suggesting a divergence at half-filling. Since we do not have access to arbitrary small doping, we cannot definitively say if there is a true divergence there, or if the divergence is eventually cut-off by some small energy/temperature scale.
The largest values occur at $\nu=0$ and $\nu+\omega=0$, i.e., transitions that involve fermions at zero energy (Fermi level). These two frequencies are connected by time-reversal symmetry, so summing over $\nu$ removes the imaginary part. This means that this divergent feature is invisible in the susceptibility. Only the magnetic channel is shown in Fig.~\ref{fig:U16:colorplots:im}, the density channel is very similar at finite $\omega$, as was already visible in Fig.~\ref{fig:U16:sumrule}.

\begin{figure}
 \includegraphics{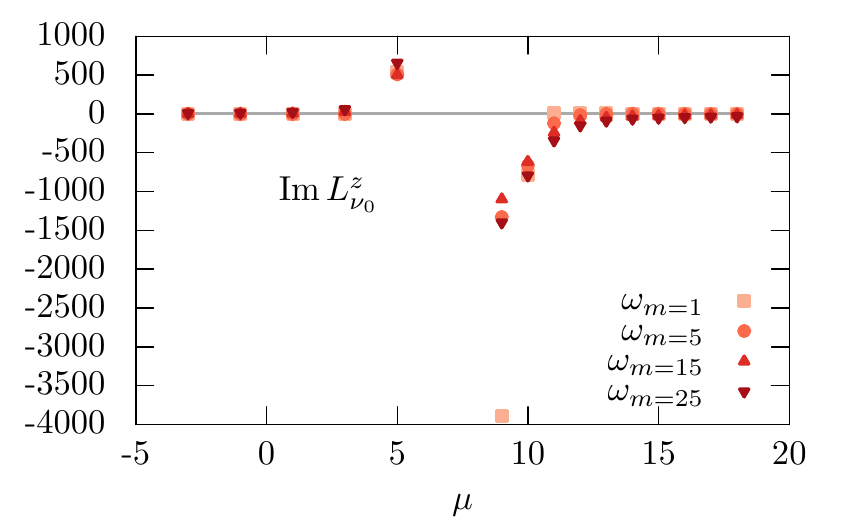}
 \caption{$\Im L^{z}_{\nu_0}$ as a function of chemical potential, for $U=16$. Half-filling occurs between $\mu=5$ and $\mu=9$, see Fig.~\ref{fig:U16:nvsmu}.}
 \label{fig:U16:muscan2}
\end{figure}

The divergence when approaching half-filling is also visible in Fig.~\ref{fig:U16:muscan2}, where the vertex at the lowest fermionic Matsubara frequency is shown as a function of the chemical potential. There is a change of sign and a large increase in magnitude at half-filling.

\begin{figure}
  \includegraphics{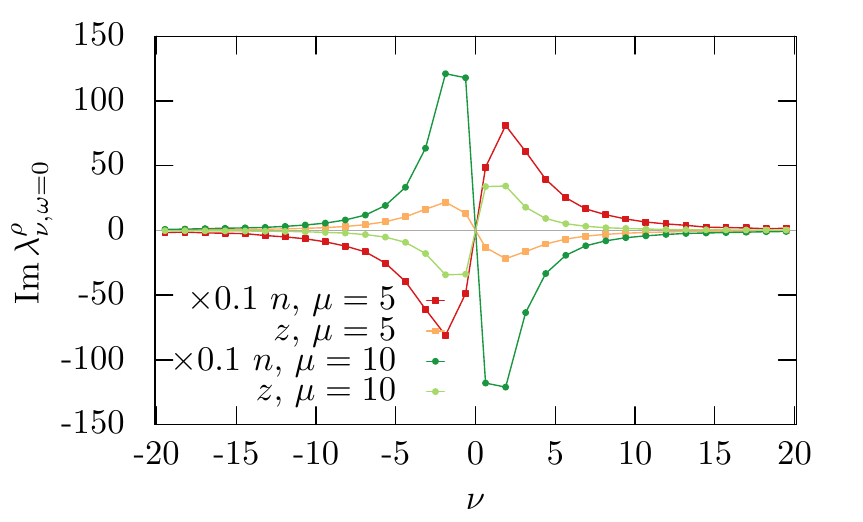}
  \caption{
  Imaginary part of the fermion-boson vertex $\lambda$ at $\omega=0$, at $U=16$, for $\mu=5$ (below half-filling) and $\mu=10$ (above half-filling). The results in the charge channel are scaled with a factor 0.1 to increase visibility.
  }
  \label{fig:U16:w0}
\end{figure}

A similar change of sign in the imaginary part occurs at zero frequency, $\omega=0$, as is visible in Fig.~\ref{fig:U16:w0}. Unlike at finite frequency, $\omega>0$, here there is still an order of magnitude difference between the charge and spin channel. The change of sign happens in both channels.

The zero frequency fermion-boson vertex of Fig.~\ref{fig:U16:w0} can be understood as the response of the  Green's function to a change in chemical potential or external magnetic field, as discussed in Appendix~\ref{app:w0}. In this case, we actually have access to the Green's function as a function of chemical potential $\mu$, so we can compare the charge channel of the vertex with the actual response to changes in $\mu$. The observation that $\lambda^{n}_{\nu,\omega=0}$ changes sign at half-filling, for any value of $\nu$, corresponds to the change of slope of $\Sigma_\nu$ as a function of $\mu$. The second observation from Fig~\ref{fig:U16:w0}, that this change of sign dramatically increases in magnitude for small $\nu$, suggests non-smooth behaviour of $\Sigma$ as a function of $\mu$. 

So, to understand the charge vertex, we can also look at the \emph{change} in self-energy as a function of chemical potential.
For the Mott insulating system, the shape of the self-energy is rather well-known.
As a function of $\mu$, $\abs{\Im \Sigma_{\nu_0}}$ reaches a maximum at half-filling (Mott insulator), corresponding to $\Im L=0$ exactly at half-filling. 
However, the value of $\abs{\Im \Sigma_{\nu_0}}$ corresponding to this maximum is divergent when the insulating phase is approached, $\Im \Sigma_{\nu_0} \sim Z^{-1}$ where $Z$ is the quasiparticle weight, so that $L$ is also divergent upon approaching the insulator.
In practice, the inverse quasiparticle weight and $\Im \Sigma$ stay finite in simulations at finite temperature, so that $L$ and $\lambda$ should also approach some finite maximum. However, since these vertices correspond to a derivative, $dZ^{-1}/d\mu$, this maximum can turn out to be extremely large.

Vertex divergences have received considerable attention recently~\cite{Schafer13,Janis14,Kozik15,Ribic16,Gunnarsson16,Schafer16,Vucicevic18,Chalupa18}. These investigations differ from our study in several ways. First of all, we find divergent behaviour as a function of chemical potential/doping at constant $U$, whereas most previous studies were performed at half-filling as a scan over $U$. In fact, this divergent behaviour shows up in the imaginary part of the vertex, which is exactly zero in the particle-hole symmetric situations that are usually studied.
Second, the divergence found here occurs in a thermodynamic response function of the system, not just in two-particle irreducible quantities~\cite{Chalupa18}. The divergence is directly related to the (change in) quasiparticle weight $Z$, which is experimentally observable. 

The fermion-boson vertex studied here is the one of the auxiliary impurity model.
Related divergent features in the momentum-resolved fermion-boson vertex of the lattice model around the metal-insulator transition have recently been found~\cite{Krien18}, in the real part at half-filling, where once again the quasiparticle weight $Z$ plays a crucial role. 

\section{Conclusion and discussion}

The fermion-boson vertex of impurity models appears in three ways in the theory of correlated fermions: as a two-particle correlation function, as the response of the Green's function to external fields and as an object in diagrammatic extensions of DMFT. In this work we have investigated the properties of this vertex. 

We started with the perturbative structure of this vertex at small $U$, following previous works~\cite{Rohringer12,HummelThesis}, which provides an intuitive understanding of the main frequency structures of the vertex. We then discussed the Ward identity for the fermion-boson vertex and used this identity to introduce an efficient description of the high-frequency behaviour of the vertex. Such an efficient description is particularly valuable with an eye on multi-band diagrammatic extensions of DMFT. It might also help to understand the simplified momentum  structure of the fermion-boson vertex found in cluster simulations~\cite{Ayral17}. We also discussed two frequency sum rules and the real energy structure of the fermion-boson vertex. 

We have used this knowledge of the fermion-boson vertex to study the doping driven metal-insulator transition. Close to the insulating phase, the charge and magnetic channel develop dramatic differences. Characteristic for the insulator is the vanishing compressibility while the magnetic susceptibility stays finite. This difference originates in the fermion-boson vertex. In particular, we find that fermions with small $\nu$ (Close to the Fermi surface) give a large \emph{negative} contribution to $dn/d\mu$. Compensation by the slowly decaying vertex at large $\nu$ ensures that $dn/d\mu$ ends up positive, as required from thermodynamic principles. 
The response to external fields with high frequency is completely different. At large Matsubara frequencies, the charge and spin response is identical~\cite{Krien17} and this is also visible in the fermion-boson vertex.

Our analysis in this work is aimed only at impurities from Dynamical Mean-Field Theory. There are extensions of DMFT that introduce retarded interactions. The impact of that type of interaction is discussed in Appendix~\ref{app:retarded}.

\acknowledgments
 
E.G.C.P. v. L. and M.I.K. acknowledge support from ERC Advanced Grant 338957 FEMTO/NANO.
A.I.L. acknowledges support from the excellence cluster ``The Hamburg Centre for Ultrafast Imaging - Structure, Dynamics and Control of Matter at the Atomic Scale''. 
The authors thank Patrik Thunstr\"om and Georg Rohringer for useful discussion about the atomic limit, and Patrick Chalupa and Alessandro Toschi for discussion of vertex divergences.
The auxiliary impurity model was solved using a modified version of the open source CT-HYB solver~\cite{Hafermann13,Hafermann14} based on the ALPS libraries~\cite{ALPS2}.

\appendix

\section{Feynman rules}
\label{app:feynmanrules}

In this appendix, we give the Feynman rules that are used to derive the diagrammatic expressions in Sec.~\ref{sec:diagrams}. 
\begin{itemize}
 \item Draw all topologically distinct connected skeleton diagrams involving propagators for up and down fermions (red/blue) and the Hubbard interaction $U$. Every interaction has one incoming and outgoing line of both colors.
 \item Amputate the incoming/outgoing fermion lines on the left-hand side.
 \item Associate a factor $g$ with every remaining fermion line.
 \item Associate a factor $U$ with every Hubbard interaction $U$.
 \item Start with prefactor $+1$, every fermion loop contributes a factor $-1$, every interaction contributes a factor $-1$. 
 \item Assign frequency variables to every fermion line, sum over internal frequencies, every internal summation carries a factor $1/\beta$. The external frequency $\nu'$ is also summed over, this summation also carries a factor $1/\beta$.
\end{itemize}

It can be useful, as discussed for diagram (a) in the main text, to replace a bubble of two Green's functions $g$ by a susceptibility $\chi$.

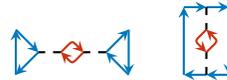
\begin{figure}
\begin{tikzpicture}
\begin{scope}[xscale=-1]
\draw[thick,->,fmBlue] (-0.3,-0.3) -- (0.0,0.0) ;
\draw[thick,<-,fmBlue] (-0.3,+0.3) -- (0.,0) ;
\draw[thick,->,fmBlue] (0.9,0) -- (1.2,-0.3) ;
\draw[thick,<-,fmBlue] (0.9,0) -- (1.2,0.3) ;
\draw[thick,->,fmRed,out=-45,in=-90-45,looseness=2] (0.3,0.) to (0.6,0) ;
\draw[thick,<-,fmRed,out=45,in=90+45,looseness=2] (0.3,0.) to (0.6,0) ;
\draw[thick,dashed] (0,0) -- (0.3,0) ;
\draw[thick,dashed] (0.6,0) -- (0.9,0) ;
\draw[thick,->,fmBlue] (-0.3,0.3) -- (-0.3,-0.3) ;
\draw[thick,<-,fmBlue] (1.2,0.3) -- (1.2,-0.3) ;
\end{scope}
\end{tikzpicture}%
\,\,\,\,\,\,\,\,\,\,\,
\begin{tikzpicture}
\begin{scope}[xscale=-1]
\draw[thick,->,fmBlue] (0,-0.6) -- (0.3,-0.6) ;
\draw[thick,<-,fmBlue] (0,+0.3) -- (0.3,0.3) ;
\draw[thick,->,fmBlue] (0.3,-0.6) -- (0.6,-0.6) ;
\draw[thick,<-,fmBlue] (0.3,0.3) -- (0.6,0.3) ;
\draw[thick,->,fmRed,out=90+45,in=-90-45,looseness=2] (0.3,-0.3) to (0.3,0) ;
\draw[thick,<-,fmRed,out=45,in=-45,looseness=2] (0.3,-0.3) to (0.3,0) ;
\draw[thick,dashed] (0.3,-0.6) -- (0.3,-0.3) ;
\draw[thick,dashed] (0.3,0.3) -- (0.3,0.0) ;
\draw[thick,->,fmBlue] (0,0.3) -- (0,-0.6) ;
\draw[thick,<-,fmBlue] (0.6,0.3) -- (0.6,-0.6) ;
\end{scope}
\end{tikzpicture}
\caption{Counting the number of fermion loops for diagrams (c) and (d) from Table~\ref{table:diagrams}. The diagram on the left has three fermion loops, the diagram on the right has two fermion loops.
}
\label{fig:loops}
\end{figure}

The concept of counting loops deserves a bit more explanation. The diagrams for $L$ have four external lines, the two fermions that form the boson and the two normal fermions. These external points have to be handled consistently to determine the overall sign of the diagram. This can be understood by comparing diagrams (c) and (d), which differ in sign exactly because of how the external lines are attached. 
To conveniently and consistently count the number of loops, we found it useful to replace the four-fermion interaction vertex by a ``mediating boson'' that couples to propagators of opposite spin on both ends, to connect the two fermionic lines that form the boson and to also connect the other two fermions. In Fig.~\ref{fig:loops} we illustrate this for diagrams (c) and (d), which differ in sign because they then have 3 and 2 loops, respectively.

\section{Bubble}
\label{app:bubble}

We introduce a shorthand notation for a convolution of two Green's function (bubbles),
\begin{align}
 B_\omega =& \frac{1}{\beta} \sum_{\nu} g_{\nu} g_{\nu+\omega}, \notag \\
 B^{pp}_{\omega} = & \frac{1}{\beta} \sum_{\nu} g_{\nu} g_{-\nu+\omega}. \label{eq:bubble:def}
\end{align}
We provide their well-known frequency inversion relations for future use  $B^{pp}_{-\omega} = \left(B^{pp}_{\omega}\right)^*$, which follows from $g_{-\nu}=\left(g_\nu\right)^*$, and $B_{-\omega}=\left(B_{\omega}\right)^*=B_\omega$:
\begin{align}
 B_{\omega} =& \frac{1}{\beta} \sum_{\nu} g_{\nu} g_{\nu+\omega} \notag \\
 =& \frac{1}{\beta} \sum_{\nu} g_{-\nu}^* g_{-\nu-\omega}^* \notag \\
 =& \frac{1}{\beta} \sum_{a=-\nu-\omega} g_{a+\omega}^* g_{a}^* \notag \\ 
 =& B_{\omega}^*.
\end{align}
Note that the definition of $B$ does not include a summation over spins. In the non-interacting system, the bubble is proportional to the susceptibility, $2B_\omega=\chi^\rho_\omega(U=0)$ for both $\rho=n$ and $\rho=S^z$.

 \section{Ward identity}
 \label{app:ward}

 Our goal is to prove, for $\omega\neq 0$,
\begin{align}
    L^{\up\up}_{\nu\omega}=-&1+\frac{\Sigma_{\nu+\omega}-\Sigma_\nu}{\imath\omega} \notag \\
    +&\frac{1}{\beta}\sum_{\nu'}\frac{\Delta_{\nu'+\omega}-\Delta_{\nu'}}{\imath\omega}\gamma^{\up\up}_{\nu\nu'\omega}g_{\nu'}g_{\nu'+\omega},
    \notag\\
    L^{\up\dn}_{\nu\omega}=    +&\frac{1}{\beta}\sum_{\nu'}\frac{\Delta_{\nu'+\omega}-\Delta_{\nu'}}{\imath\omega}\gamma^{\up\dn}_{\nu\nu'\omega}g_{\nu'}g_{\nu'+\omega},\notag 
\end{align}
 
 In this case, it is more convenient to do the derivation in terms of $\lambda$ than in terms of $L$, i.e. to use the charge and spin channel explicitly.
 We start from Eq.~E3 in Ref.~\onlinecite{Krien17}. In the present case, it reads
 \begin{align}
  g_{\nu+\omega} - g_{\nu} = \frac{1}{\beta} \sum_{\nu'} \chi^{\rho}_{\nu\nu'\omega} (\Delta_{\nu'+\omega} - \Delta_{\nu'} - \imath\omega) \label{eq:ward:app:1}
 \end{align}
 where the three-frequency object $\chi^{\rho}_{\nu\nu'\omega}=g^{(4),\rho}_{\nu\nu'\omega}+2\beta g_{\nu}g_{\nu'} \delta_{\omega}\delta_{\rho,\text{c}}$ is the generalized susceptibility given in the charge/spin channel, the last term only contributes in the charge channel and the factor 2 comes from the sum over spin. 
 The generalized susceptibility is related to the fermion-boson vertex as $g_{\nu}g_{\nu+\omega}\chi_{\omega} \lambda_{\nu\omega} = -\frac{1}{\beta} \sum_{\nu'} \chi_{\nu\nu'\omega}$, for $\omega\neq 0$. The last term in the right-hand side of Eq.~\eqref{eq:ward:app:1} is independent of $\nu'$ so that the frequency sum can be performed, 
 \begin{widetext}
\begin{align}
  g_{\nu+\omega} - g_{\nu} =&  \imath\omega g_{\nu}g_{\nu+\omega}\chi^{\rho}_{\omega} \lambda^{\rho}_{\nu\omega} +  \frac{1}{\beta} \sum_{\nu'} \chi^{\rho}_{\nu\nu'\omega} (\Delta_{\nu'+\omega} - \Delta_{\nu'}) \notag \\
  g^{-1}_{\nu} - g^{-1}_{\nu+\omega} =&  \imath\omega \chi^{\rho}_{\omega}\lambda^{\rho}_{\nu\omega} +  g^{-1}_{\nu}g^{-1}_{\nu+\omega}\frac{1}{\beta} \sum_{\nu'} \chi^{\rho}_{\nu\nu'\omega} (\Delta_{\nu'+\omega} - \Delta_{\nu'}) \notag \\  
  \chi^{\rho}_{\omega}\lambda^{\rho}_{\nu\omega} =& \frac{  g^{-1}_{\nu}-g^{-1}_{\nu+\omega} }{\imath\omega} - g^{-1}_{\nu}g^{-1}_{\nu+\omega}\frac{1}{\beta} \sum_{\nu'} \chi^{\rho}_{\nu\nu'\omega} \frac{\Delta_{\nu'+\omega} - \Delta_{\nu'}}{\imath\omega} \notag \\
  \chi^{\rho}_{\omega}\lambda^{\rho}_{\nu\omega} =& \frac{  -\imath\omega + \Sigma_{\nu+\omega}-\Sigma_{\nu}+\Delta_{\nu+\omega}-\Delta_{\nu} }{\imath\omega} - g^{-1}_{\nu}g^{-1}_{\nu+\omega}\frac{1}{\beta} \sum_{\nu'} \chi^{\rho}_{\nu\nu'\omega} \frac{\Delta_{\nu'+\omega} - \Delta_{\nu'}}{\imath\omega} \notag \\  
  =&-1+\frac{\Sigma_{\nu+\omega}-\Sigma_\nu}{\imath\omega} +\frac{1}{\beta}\sum_{\nu'}\frac{\Delta_{\nu'+\omega}-\Delta_{\nu'}}{\imath\omega}\gamma^{\rho}_{\nu\nu'\omega}g_{\nu'}g_{\nu'+\omega},
  \label{eq:ward:app:2} 
 \end{align}
\end{widetext}

 The Ward identities in terms of $L$ are then recovered by simply taking the average and difference of these equations for the charge and $S^z$ channel.
 
 To obtain the Ward sum-rule, an additional sum over $\nu$ is performed. It is most convenient to start again at Eq.~\eqref{eq:ward:app:1}, where it is clear that the left-hand side vanishes after summing over $\nu$. 
 \begin{align}
 0=& \frac{1}{\beta^2} \sum_{\nu\nu'} \chi^\rho_{\nu\nu'\omega} (\Delta_{\nu'+\omega} - \Delta_{\nu'} - \imath\omega) \\
 =& -\imath\omega \chi^\rho_{\omega} - \frac{1}{\beta} \sum_{\sigma\nu'} \left(\Delta_{\nu'+\omega}-\Delta_{\nu'}\right)
 g_{\nu'}g_{\nu'+\omega}\chi^\rho_{\omega} \lambda_{\nu'\omega}, \notag
 \end{align}
 which for a paramagnetic system can be rewritten as
 \begin{align}
  -\frac{2}{\beta} \sum_{\nu} \frac{\Delta_{\nu+\omega}-\Delta_\nu}{\imath\omega} g_{\nu} g_{\nu+\omega} \lambda^\rho_{\nu\omega} = 1.
 \end{align}
 Note that here we used again the relation between generalized susceptibility and fermion-boson vertex, but this time for the sum over the first fermionic frequency.

\section{The vertex at large frequencies}
\label{app:asymptote}

Diagrammatic analysis can be used to determine the asymptotic structure of vertices~\cite{Kunes11,HummelThesis}. The main point is that if one of the external lines carries a large frequency, this same frequency is also present in the internal lines of the diagram. Since the Green's function decays as a function of frequency, this means that those diagrams also decay in magnitude. In higher order diagrams, this large frequency is usually carried by several propagators and these diagrams therefore decays more quickly. Only specific higher-order diagrams remain relevant and these can be resummed in terms of susceptibilities. 

Here, we follow a similar approach, but with an additional trick. Instead of performing a diagrammatic analysis of $L$ directly, we use the Ward identity Eq.~\eqref{eq:wardid} and plug in the asymptotic analysis of $\gamma$~\cite{HummelThesis}. The advantage of this approach is that for large $\omega$ the self-energy term already contains a decent part of the frequency structure in $\nu$, as is visible in the bottom panel of Fig.~\ref{fig:U10:plots}, and that the remaining contribution from the fermion-fermion vertex appears in a combination with single-particle quantities that enhance the convergence of the frequency sums. The resulting expressions allow for an efficient description of the fermion-boson vertex at larger frequencies, so that the exact calculations only have to be done for a small frequency window~\cite{Kunes11,Wentzell16,Tagliavini18}

Our goal is to find expressions for the vertex that only involve single-frequency objects. We start with the asymptotic expressions for the fermion-fermion vertex~\cite{HummelThesis}:
\begin{align*}
 \gamma^{\up\up}_{\nu\nu'\omega} =& 0 -U^2 \chi^{\dn\dn}_\omega  +U^2 \chi^{\dn\dn}_{\nu'-\nu} \\
 \gamma^{\up\dn}_{\nu\nu'\omega} =& -U +U^2 \chi^{\dn\dn}_{\nu'-\nu} +U^2 B^{pp}_{\nu+\nu'+\omega}.
\end{align*}
Here, the treatment of the particle-particle susceptibility differs from the particle-hole susceptibilities. The latter can be measured efficiently inside the impurity solver, so that we simply use the full susceptibility. For the particle-particle susceptibility we do not have the exact impurity result and we instead use the bubble $B^{pp} = - \chi^{pp}$.

\begin{figure}
 \includegraphics{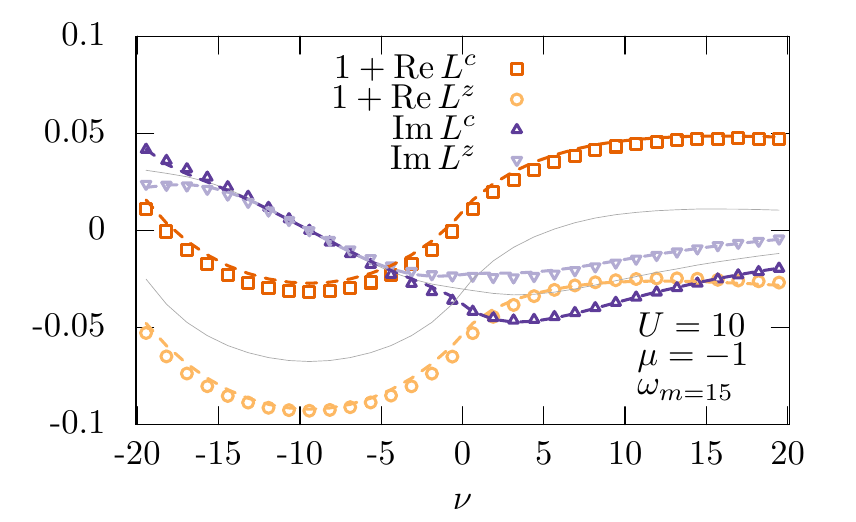}
 \includegraphics{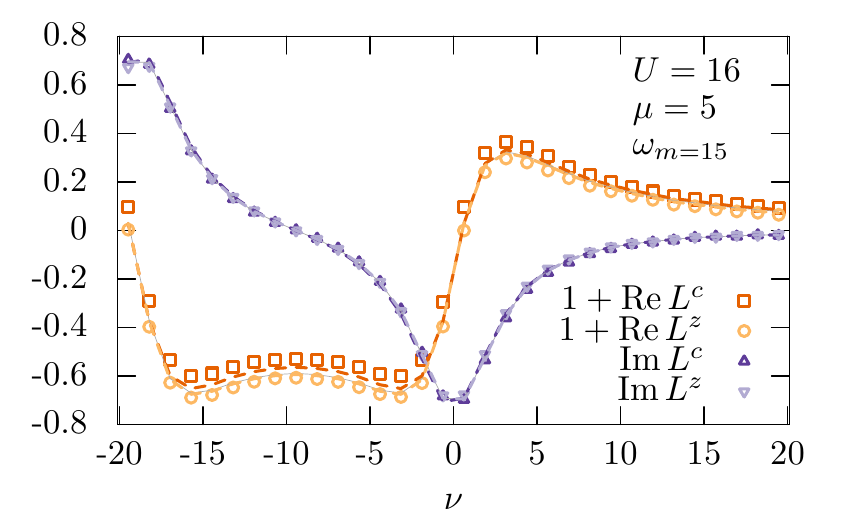}
 \caption{The fermion-boson vertex at $U=10$ and $U=16$, with $\mu=-1$ and $\mu=5$ respectively, $\beta=5$ given in the charge and spin channel, the dashed lines correspond to the asymptotic expressions of Eq.~\eqref{eq:ward:asymptote}. The gray lines are the approximation that ignores the fermion-fermion vertex completely, Eq.~\eqref{eq:ward:approx}. See also Fig.~\ref{fig:U10:plots}.}
 \label{fig:U10:asymp}
\end{figure}

Then, we insert these expressions into the Ward identity, Eq.~\eqref{eq:wardid}. In this way, we obtain expressions for $L$ that only involve $g_\nu$, $\sigma_\nu$, $\Delta_\nu$ and $\chi_\omega$.
\begin{widetext}
\begin{align}
 L^{\up\up}_{\nu\omega} =& -1 + \frac{\Sigma_{\nu+\omega}-\Sigma_\nu}{\imath\omega} + U^2 \frac{1}{\beta} \sum_{\nu'} g_{\nu'}g_{\nu'+\omega} \left(\chi^{\dn\dn}_{\nu'-\nu}-\chi^{\dn\dn}_{\omega}\right) \frac{\Delta_{\nu+\omega}-\Delta_\nu}{\imath\omega} \\
 L^{\up\dn}_{\nu\omega} =& \frac{U}{\beta} \sum_{\nu'} g_{\nu'}g_{\nu'+\omega} \left(-1+U\chi^{\dn\dn}_{\nu'-\nu}+U B^{pp}_{\nu+\nu'+\omega}\right) \frac{\Delta_{\nu+\omega}-\Delta_\nu}{\imath\omega} \label{eq:ward:asymptote}
\end{align} 
\end{widetext}
These formulas are illustrated in Fig.~\ref{fig:U10:asymp}. This figure shows that this approximation works well both qualitatively and quantitatively at these parameters. The simpler approximation of Eq.~\ref{eq:ward:approx} (gray lines) completely ignores the fermion-fermion vertex and therefore does not distinguish between the charge and spin channel. In general, we find that Eq.~\ref{eq:ward:asymptote} works well at large $\omega$, even for strong interaction strengths. Close to the metal-insulator transition, deviations are a bit larger but still very reasonable.

 \section{Non-interacting system and Ward identity}
\label{app:deltaU0} 

In DMFT for a non-interacting system, $\Sigma=0$, so $g^{-1}_\nu = i\nu - \Delta_{\nu}$. We can use this to simplify the expression that occurs in the Ward identity for the non-interacting system:
\begin{align*}
 \frac{\Delta_{\nu'+\omega} -\Delta_{\nu'} }{\imath\omega} g_{\nu'} g_{\nu'+\omega}= g_{\nu'} g_{\nu'+\omega} + \frac{1}{\imath\omega} (g_{\nu'} - g_{\nu'+\omega} ) \\
\sum_{\nu'} \frac{\Delta_{\nu'+\omega} -\Delta_{\nu'} }{\imath\omega} g_{\nu'} g_{\nu'+\omega}= \sum_{\nu'} g_{\nu'} g_{\nu'+\omega},
\end{align*}
where the second term on the right-hand side summed to zero as a telescoping series.

For small $U$, to leading order $g$ is still given by the non-interacting Green's function and $\gamma^{\up\dn}\propto U$ is independent of frequency and can be taken out of frequency sums and $\gamma^{\up\up}$ is of order $U^2$ and can be neglected. Inserting the previous result into the Ward identity Eq.~\eqref{eq:wardid}, we indeed recover the result of diagram (a).

\section{Fermion-boson vertex at zero frequency}
\label{app:w0}

We have derived the Ward identity for the fermion-boson vertex $L_{\nu\omega}$ for $\omega \neq 0$. It is also possible to derive an insightful relation for the vertex at $\omega=0$. 
The zero frequency fermion-boson vertex can be understood as the response of the  Green's function to a change in chemical potential or external magnetic field~\footnote{Again, we remind the reader of the fact that the impurity vertex describes the response at constant $\Delta$, as discussed in footnote~\onlinecite{endnote108}.}. 
Going from $\lambda$, or $L$ to the response, we have to remember the additional factor $g g$ in Eq.~\eqref{eq:L:abstract2}, meaning that our observations do not translate directly. In fact, we can rewrite the equations in a more natural way, by using the relation between self-energy $\Sigma$ and Green's function $g$,
\begin{align}
 g_\nu^{-1} =& \imath \nu - \Delta_\nu +\mu -\Sigma_\nu \notag \\
 \frac{dg_\nu}{d\mu} =& -g_\nu^2 \frac{dg_\nu^{-1}}{d\mu} \notag \\
 =& g_\nu^2 \left(-1+\frac{d\Sigma_\nu}{d\mu} \right) \notag \\
 L^n_{\nu,\omega=0} =& -1+\frac{d\Sigma_\nu}{d\mu}.
\end{align}
Here, we used that the fermion-boson vertex corresponds to changing $\mu$ while keeping $\Delta$ constant. 
This relation can be seen as the zero-frequency equivalent of the Ward identity~\cite{Krien18}, Eq.~\eqref{eq:wardid}. In the derivation of Ref.~\onlinecite{Krien18}, this relation is shown to hold even when only approximations of $\Sigma$ and $L$ are known, as long as those approximations satisfy the Ward identity. Our derivation here only applies to models with an exact solution, but that is sufficient for our purposes here.

According to Fig.~\ref{fig:U16:w0}, the change in sign of the fermion-boson vertex occurs at all frequencies $\nu$, including the large frequency limit. This limit can be understood in terms of the asymptotics of $\Sigma$, see e.g,~Ref.~\onlinecite{Rohringer16},
\begin{align}
 \Sigma_\nu \overset{\nu\rightarrow\infty}{=} \frac{U\av{n}}{2} +U^2 \frac{\av{n}}{2}\left(1-\frac{\av{n}}{2}\right)\frac{1}{i\nu}+\ldots.
\end{align}
By taking the derivative with respect to $\mu$, we find that the imaginary part of the fermion-boson vertex decays as $\frac{1}{i\nu}$ and is proportional to the compressibility, to $U^2$ and to the deviation from half-filling,
\begin{align}
 \Im L^n_{\nu,\omega=0}=\Im \frac{d\Sigma_{\nu}}{d\mu} \overset{\nu\rightarrow\infty}{=} -\frac{U^2}{2\nu} \left(1-\av{n}\right) \frac{d\av{n}}{d\mu} . \label{eq:L:sigma:asymptote}
\end{align}
The $U^2$ factor is consistent with our observation that lowest-order diagam (a) is real.
Since both $U^2$ and $dn/d\mu$ are positive, the sign change at half-filling is clear in this expression, and occurs regardless of interaction strength.
Going from $L$ to $\lambda$, we simply need to divide by the susceptibility at $\omega=0$, i.e., by $-d\av{n}/d\mu$,
\begin{align}
 \Im \lambda^{n}_{\nu,\omega=0} \overset{\nu\rightarrow\infty}{=} \frac{U^2}{2\nu} \left(1-\av{n}\right).
\end{align}

\section{Retarded interaction}
\label{app:retarded}

Several extensions of DMFT, such as EDMFT\cite{Sengupta95,Si96,Kajueter96,Smith00,Chitra00,Chitra01}, EDMFT+$GW$~\cite{Sun02,Sun03,Biermann03}, dual boson~\cite{Rubtsov12} and TRILEX~\cite{Ayral15,Ayral16,Vucicevic17,Ayral17}, use impurity models with retarded (dynamical) interactions. This means that the impurity action, Eq.~\eqref{eq:Simp}, contains terms like $U_\omega n_\omega n_\omega$ or $U_\omega S^z_\omega S^z_\omega$.
Similar retarded interactions occur in effective Hubbard models obtained by constrained RPA~\cite{Aryasetiawan04}.
To what extent do the results obtained in this manuscript generalize to models with dynamical interactions?

The diagrammatic analysis of Sec.~\ref{sec:diagrams} can proceed in a rather similar way, taking into account that the interaction $U$ needs to be replaced by $U^{\sigma\sigma'}(\omega_i)$, with $\omega_i$ the appropriate transferred frequency. In general, the Hubbard interaction now also couples two fermion lines with the same spin (albeit at different times). This means, for example, that there will be a counterpart to diagram (a) for $L^{\up\up}$.

The Ward identity is based on the continuity equation and contains the commutation relation $\left[ \rho, H_{\text{AIM}} \right]$. As long as the retarded interaction commutes with the bosonic variable $\rho$ under investigation, the Ward identity will stay the same. However, when they do not commute, additional terms in the Ward identity will appear~\cite{Krien17}. For example, additional terms will appear for the fermion-boson vertex $L^z$ in an impurity model with a retarded interaction in the $S^x$ channel. This also means that the Ward sum rule, Eq.~\eqref{eq:sumruleward}, will contain additional terms. On the other hand, the sum rule Eq.~\eqref{eq:sumrule1} still holds since it follows from the definition of the correlation functions. 

 \bibliography{references}
 
\end{document}